\documentclass[sigconf]{acmart}

\usepackage{wrapfig}
\usepackage{array}
\usepackage{circledsteps} 
\usepackage{listings}
\usepackage{microtype}
\usepackage{paralist, enumitem}
\usepackage{subcaption}
\usepackage{tcolorbox}
\usepackage{xcolor}
\usepackage{xspace}
\usepackage{multirow}

\lstdefinestyle{fJava}{
  language=Java,
  basicstyle=\ttfamily\scriptsize,
  keywordstyle=\color{purple},
  commentstyle=\color{blue},
  stringstyle=\color{red},
  numbers=left,
  numberstyle=\tiny,
  numbersep=5pt,
  breaklines=true,
  breakatwhitespace=true,
  columns=flexible,
  tabsize=1,
  showstringspaces=false,
  escapeinside={(*@}{@*)},
}

\lstdefinestyle{iJava}{
  language=Java,
  basicstyle=\ttfamily\small,
  keywordstyle=\color{purple},
  commentstyle=\color{blue},
  stringstyle=\color{red},
  numbersep=5pt,
  breaklines=true,
  breakatwhitespace=true,
  columns=flexible,
  tabsize=1,
  showstringspaces=false,
  escapeinside={(*@}{@*)},
  mathescape
}

\lstdefinestyle{fSQL}{
  language=SQL,
  basicstyle=\ttfamily\scriptsize,
  keywordstyle=\color{purple},
  commentstyle=\color{blue},
  stringstyle=\color{red},
  breaklines=true,
  breakatwhitespace=true,
  columns=flexible,
  tabsize=1,
  showstringspaces=false,
  escapeinside={(*@}{@*)},
}

\newcommand{\mathlst}[1]{\text{\lstinline|#1|}}

\newcommand{\Tool}{\textsc{Merlin}}

\newcommand{\sadra}[1]{{\color{orange!70!black} [#1]$_\text{Sadra}$}}

\renewcommand{\sadra}[1]{}

\newcommand{\mclearpage}{}
\newcommand{\mmclearpage}{}

\title{Generating Complex Code Analyzers from Natural Language Questions}

\author{Amirmohammad Nazari}
\orcid{}
\affiliation{
  \institution{University of Southern California}
  \city{Los Angeles}
  \state{CA}
  \country{USA}
}
\email{nazaria@usc.edu}

\author{Sadra Sabouri}
\orcid{}
\affiliation{
  \institution{University of Southern California}
  \city{Los Angeles}
  \state{CA}
  \country{USA}
}
\email{sabourih@usc.edu}

\author{Wang Bill Zhu}
\orcid{}
\affiliation{
  \institution{University of Southern California}
  \city{Los Angeles}
  \state{CA}
  \country{USA}
}
\email{wangzhu@usc.edu}

\author{Robin Jia}
\orcid{}
\affiliation{
  \institution{University of Southern California}
  \city{Los Angeles}
  \state{CA}
  \country{USA}
}
\email{robinjia@usc.edu}

\author{Souti Chattopadhyay}
\orcid{}
\affiliation{
  \institution{University of Southern California}
  \city{Los Angeles}
  \state{CA}
  \country{USA}
}
\email{schattop@usc.edu}

\author{Mukund Raghothaman}
\orcid{}
\affiliation{
  \institution{University of Southern California}
  \city{Los Angeles}
  \state{CA}
  \country{USA}
}
\email{raghotha@usc.edu}

\begin{document}

\begin{abstract}
Many software development tasks, such as implementing features and fixing bugs, begin with developers posing questions about a codebase. However, answering questions about codebases that span millions of lines of code across thousands of files is non-trivial. Standard tools like \texttt{grep} cannot answer questions requiring semantic or inter-procedural reasoning, and large language models (LLMs) struggle with large codebases due to resource and context constraints.
In this paper, we present \Tool, a new system for answering free-form questions that require analytical reasoning about code. \Tool{} integrates an LLM with CodeQL, a program analysis framework that supports expressive queries over large codebases.
We face two principal challenges in the design of such systems:
First, program analysis queries are diverse and semantically complex; as a result, even syntactically well-formed queries frequently produce degenerate/empty results.
Furthermore, relatively few CodeQL queries are available online, limiting the out-of-the-box effectiveness of LLMs as CodeQL query generators.
We address these challenges by developing a RAG-based iterative query-generation approach and a novel self-test technique. Our query debugging technique builds on the idea of assistive queries, which generate concrete witnesses that expose and explain semantic flaws in candidate queries.
We evaluate \Tool{} through both experimental and user studies.
Over a set of natural language questions derived from common bug-finding tasks, \Tool{} discovered not only the majority of software issues reported by other approaches, but also issues that would have otherwise remained undetected.
Through a within-subject user study, we found that access to \Tool{} increased task accuracy by an average of 3.8\(\times\) and simultaneously reduced the time for programmers to complete all tasks by 31\%.
\end{abstract}

\maketitle

\mclearpage \section{Introduction}
\label{sec:intro}


Asking questions about code is foundational to the software engineering process~%
\cite{ko2006exploring, sillito2008asking}. Developers routinely query their codebases throughout the
entire software lifecycle, from initial onboarding and active feature development to post-deployment
maintenance~\cite{latoza2010hard, ko2007information}. However, obtaining accurate answers to these
questions is a significant bottleneck, known to consume a substantial fraction of development time
and effort, even among experts~\cite{latoza2010hard, minelli2015know}.

Among the many questions, a particularly challenging class consists of \emph{deep analytical
questions} that require reasoning over the global structure and semantics of a codebase. Examples
include identifying all locations where a specific API is used, finding methods that are overloaded
in particular ways, or locating calls to a method where arguments satisfy specific semantic
constraints. Unlike local or syntactic questions, these queries cannot be answered by inspecting a
single file or code fragment in isolation. Instead, they require combining information across
multiple program elements, such as types, control flow, and call relationships.

Common programming tools offer limited support for answering these analytical questions.
Developers often rely on ad hoc utilities such as \texttt{grep} or IDE features to
incrementally explore a codebase. While these can surface individual code locations, they do
not provide a mechanism for expressing complex queries or systematically combining partial
results into a coherent answer.

Alternatively, programmers may turn to large language models (LLMs), which provide a natural
interface for posing free-form questions about code. However, applying LLMs to large, real-world
codebases remain challenging. Current LLM-based tools typically operate over plain text
representations of code and lack a structured, persistent view of the codebase.

As a result, answering questions that require reasoning over large codebases is often unreliable or
prohibitively expensive, as relevant context may exceed the model's context window or be only
partially represented~\cite{ding2024longrope, an2024make}.



To address these limitations, we introduce \Tool, a natural language question answering system
designed to support analytical queries over large codebases. \Tool{} combines the convenience of a
natural language interface with CodeQL~\cite{codeql}, a declarative program analysis framework that
provides a structured representation of a codebase and supports a wide range of static analyses.

CodeQL models a codebase as a relational database containing syntactic and semantic information
about the program, including types, control-flow and data-flow relationships, and other program
properties. Analyses are then expressed using a SQL-like language, enabling precise reasoning about
program structure and behavior at scale.

\Tool{} uses an LLM to translate a developer's natural language question into a corresponding CodeQL
query, which is then executed over the codebase. The resulting set of code locations and program
elements is returned to the user, grounding answers in the results of program analysis rather than
in the LLM's internal textual representation of the codebase.


Developing \Tool{} required addressing two technical challenges. First, the availability of CodeQL
queries are limited, which makes it difficult for pre-trained LLMs to reliably generate well-formed
analysis queries or to fine-tune open-weight models. To address this, \Tool{} combines the LLM with
CodeQL documentation and compiler feedback in an iterative, RAG-like loop to ensure syntactic
correctness.

Second, even syntactically valid CodeQL queries may fail to capture the programmer's intended
semantics, as closely related queries can differ subtly in meaning. We address this challenge using
a \emph{self-test} mechanism in which the LLM generates code examples that should or should not
satisfy the query, allowing us to detect degenerate queries before applying them to the codebase.

While self-tests are effective at detecting semantic mismatches, they do not explain why a query
fails or how it should be corrected. To support diagnosis and refinement, we introduce
\emph{assistive queries}, which allow the LLM to generate auxiliary CodeQL queries that expose
intermediate program properties relevant to the original question. These assistive queries help
illuminate the causes of semantic errors, functioning analogously to diagnostic print statements
during debugging.



In order to evaluate the \Tool{} system, we started by developing a set of benchmarks: We observed
that bug-finding problems (e.g., finding calls to suspiciously named methods such as Java's
\lstinline|Array.equals()|, calls to non-\lstinline|final| methods in object constructors, or direct
manipulation of string data possibly leading to code injection vulnerabilities) are a rich source of
analytical questions about program behavior and structure. We constructed a set of benchmarks by
consulting the analyses performed by SpotBugs~\cite{SpotBugs, findbugs} and GitHub Security Lab~%
\cite{GitHubSecurityLab}, and phrasing the underlying problems as natural language questions that
programmers might naturally ask.

We found that \Tool{} is not only effective at reproducing responses produced by these baseline
tools, but also flags many code snippets that were not identified by the baselines. To understand
why \Tool{} finds so many additional code snippets, we conducted a survey in which programmers
annotate identified code snippets as relevant or irrelevant. This survey confirmed that \Tool{}
finds genuinely relevant code snippets that are overlooked by other tools. In fact, when taking
human relevance judgments as ground truth, \Tool{} achieves both higher precision and higher recall
than SpotBugs.

Finally, we conducted an end-to-end user study to determine whether \Tool{} can truly assist
programmers in realistic software engineering tasks. We started by presenting programmers with
guidelines from the SEI CERT coding standards~\cite{CertJava}, and asking them to find and fix all
violations in a given codebase. As before, identifying violations involved answering some analytical
question about the codebase. One group had access to \Tool, while the control group did not (but
could use any other tool, including AI-based tools). Having access to \Tool{} increased task
accuracy by an average of $3.8\times$ and simultaneously reduced time to complete all tasks by 31\%.

\paragraph{Contributions}
To summarize, in this paper, we:
\begin{enumerate}
\item Present \Tool, a system that integrates large language models with program analyzers to
  effectively and reliably answer free-form natural language questions about large codebases.
\item Develop a RAG-based iterative procedure and a self-test technique combined with the idea of
  assistive queries to reliably generate complex program analysis queries, despite the
  unavailability of large query corpora.
\item Conduct an empirical evaluation, a survey, and a user study which show that \Tool{}
  answers realistic developer questions with high precision and coverage.
\end{enumerate}

\mclearpage \section{Motivation and Interaction Mechanism}

\label{sec:motiv}

\begin{figure*}
\centering
\begin{subfigure}[b]{.3\textwidth}
  \centering
  \includegraphics[width=\columnwidth]{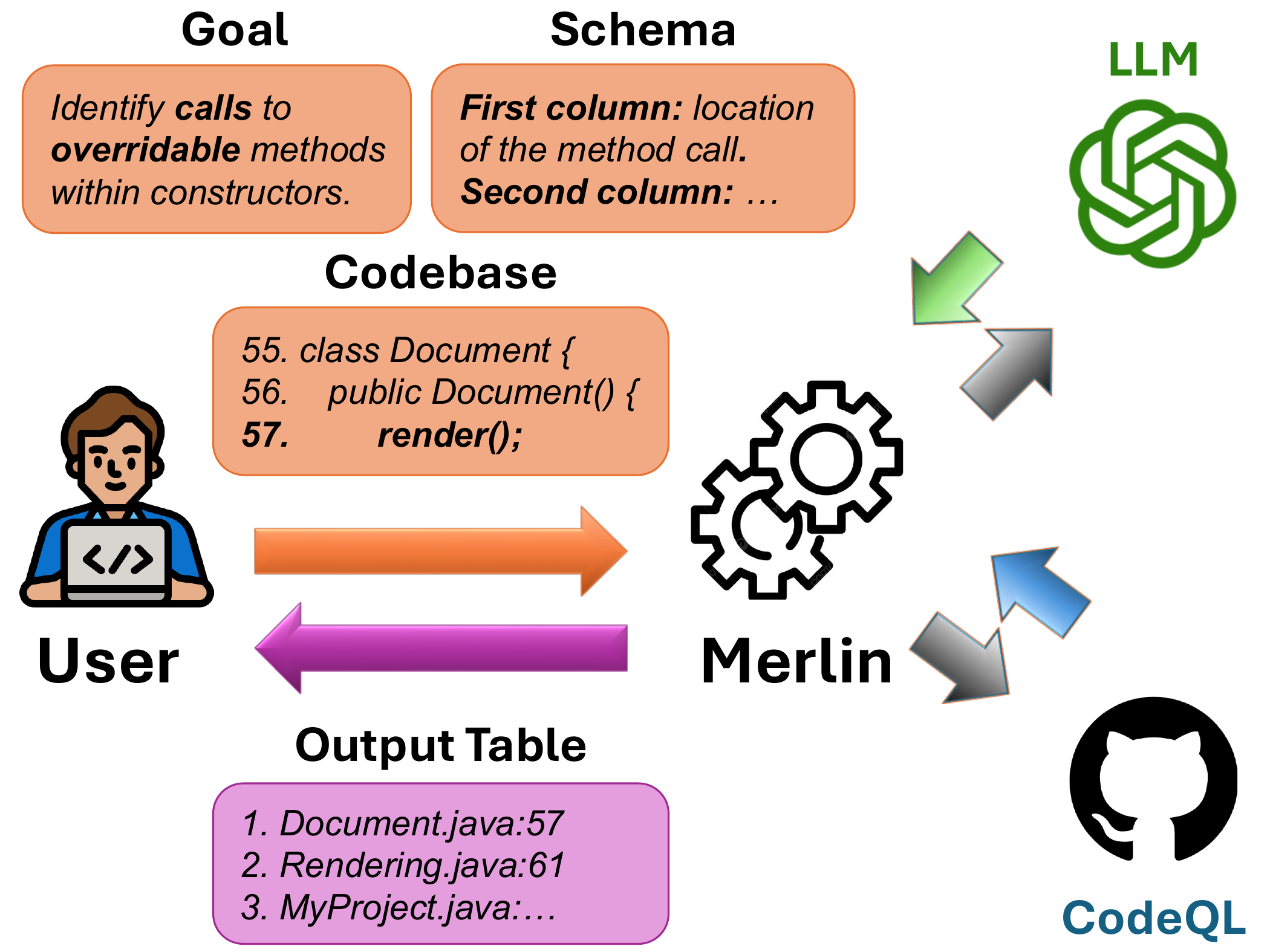}
  \caption{}
  \label{sfig:motiv:interface}
\end{subfigure}
\hspace{.06\textwidth}
\begin{subfigure}[b]{.24\textwidth}
  \centering
  \begin{lstlisting}[style=fJava, basicstyle=\tiny\ttfamily]
class SuperClass {
  public SuperClass () {
    doLogic();
  }
  public void doLogic() {
    System.out.println("Superclass!");
  }
}
class SubClass extends SuperClass {
  private String color = "red";
  public void doLogic() {
    System.out.println("Subclass! Color: " + color);
  }
}
  \end{lstlisting}
  \caption{}
  \label{sfig:motiv:met05-j-ex}
\end{subfigure}
\hspace{.04\textwidth}
\begin{subfigure}[b]{.33\textwidth}
  \centering
  \begin{lstlisting}[style=fSQL, basicstyle=\tiny\ttfamily]
import java

from Constructor c, MethodCall mc, Method m
where
  mc.getEnclosingCallable() = c and
  m = mc.getMethod() and
  m.isOverridable()
select mc.getLocation(), m.getLocation()
  \end{lstlisting}
  \caption{}
  \label{sfig:motiv:met05-j-ql}
\end{subfigure}
\caption{The \Tool{} user interaction model (\ref{sfig:motiv:interface}).
  Example Java program in which a constructor calls an overridable method
  (\ref{sfig:motiv:met05-j-ex}). The concern is that the \lstinline|subclass.doLogic()|
  method might access the \lstinline|subclass.color| variable before it was initialized.
  Automatically generated CodeQL query to find violations of the MET05-J coding guideline
  (\ref{sfig:motiv:met05-j-ql}). Notice the need to access various kinds of information about the
  program, such as whether methods are overridable and to navigate between method calls and
  the enclosing function, and to relate these pieces of information using Boolean connectives.}
\label{fig:motiv}
\end{figure*}

Consider the following coding guideline included as part of the CERT Oracle Coding Standard for
Java, a library of guidelines for writing secure Java code~\cite{MET05J, CertJava}:
\guillemotleft{}\emph{Ensure that constructors do not call overridable methods.}\guillemotright{}
The goal is to forbid patterns of code such as that shown in Figure~\ref{sfig:motiv:met05-j-ex}: If
the constructor of a class $C$ invokes a method $m$, and the class $C$ is subsequently extended into
a subclass $C'$ with an overridden implementation of $m$, then the original class $C$ might invoke
code operating on the subclass before the subclass has had an opportunity to fully initialize
itself.

After becoming aware of this guideline, the programmer might wish to search their codebase for
instances of its violation. Today, they might approach this task in one of three ways: hand
inspection, using tools such as \texttt{grep} and relying on various forms of IDE support, or by
consulting an AI coding assistant such as Claude Code~\cite{claude-code}, Copilot~\cite{Copilot}, or
Cursor~\cite{cursor}.
Hand inspection is infeasible for all but the smallest codebases.
Tools such as \texttt{grep} are focused on the task of finding textual patterns and are unsuited for
this task.
Finally, there are several challenges in using contemporary LLMs for this type of analysis:
\begin{enumerate}
\item 
  The quality of the output is inconsistent. Given the above question and the codebase from our user
  study, Gemini 3 Pro reports eight locations, all of which turn out to be
  false positives. In fact, there are only four locations in our codebase that violate the
  guideline. In other situations, the LLM frequently flags empty or irrelevant lines of code.
\item 
  LLMs and AI coding assistants need large amounts of context to answer such questions. At the same
  time, they impose resource-use limits, such as on the number of uploaded files. Specifically,
  Gemini 3 Pro limits uploads to 1000 files. This necessitates a time-consuming partitioning of the
  codebase into smaller subsets. In addition, this might also lead to errors, as semantically
  related files may be accidentally separated across partitions.
\item 
  LLM-based analysis suffers from high latency and occasional non-termination. Queries can require
  substantial execution time and may return no result, necessitating repeated query attempts. These
  issues are particularly pronounced on large codebases and were frequently observed in our user
  study.
\end{enumerate}
Together, these observations illustrate the limitations of purely LLM-based tools in reliably
performing precise and exhaustive program reasoning tasks.


Answering this question fundamentally involves analytical reasoning about the codebase:
one has to find all class constructors, identify the methods invoked, and determine whether any of
them are overridable.
AI agents such as Claude Code are a natural first candidate for tackling this kind of complex
reasoning. Given the question and the codebase, Claude Code performs step-by-step reasoning by
sequentially invoking external tools such as \texttt{grep} and then aggregating information from
their outputs to produce an answer. However, in our experiments, Claude Code (nondeterministically)
reports only two locations, neither of which corresponds to any of the four correct locations.%


Observe also that this question---``\emph{Which constructors in my codebase call overridable
methods?}''---is just one example of a wide variety of questions that are routinely asked by
software developers. Asking and answering such questions is often the first part of many software
development processes, where engineers gather information about their codebase, and make plans for
how to fix bugs and add features. Our system, \Tool, is an LLM-backed program analysis system which
allows the user to answer a variety of such questions about their code.

\begin{figure*}
\centering
\includegraphics[width=\linewidth]{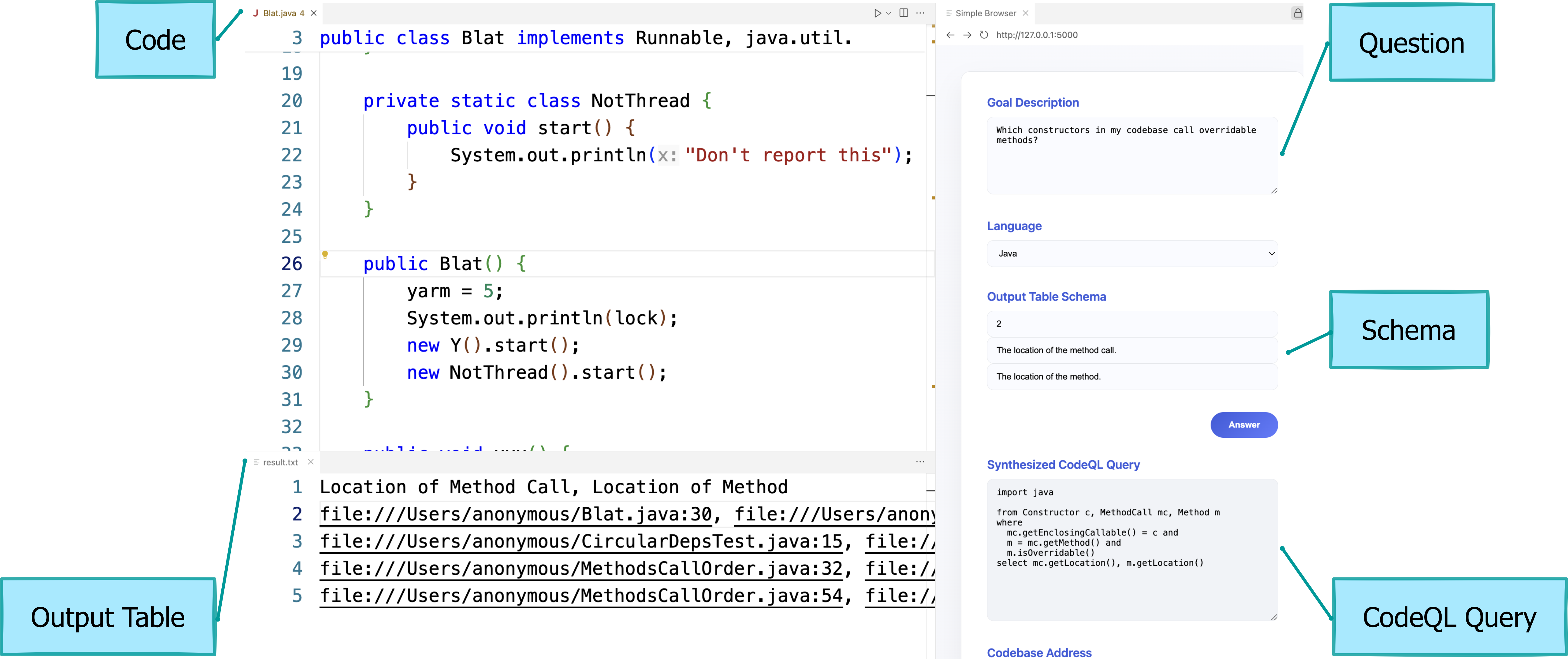}
\caption{The \Tool{} user interface. While working with a codebase, the user specifies a high-level
  question along with the desired output table schema. In response, \Tool{} returns the resulting
  output table together with the corresponding CodeQL query.}
\label{fig:motiv:ui}
\end{figure*}

We describe the high-level interaction mechanism between \Tool{} and the user in Figure~%
\ref{sfig:motiv:interface}, and a screenshot of its interface in Figure~\ref{fig:motiv:ui}. The user
opens the codebase in their IDE, and provides a textual description of their question and the
desired schema for the response. The schema specifies the desired number of columns in the output
table and a description of the expected contents of each column.%

Under the hood, \Tool{} integrates the LLM with CodeQL, a customizable program analysis tool~%
\cite{codeql}. CodeQL exports a view of the codebase as a database which contains various forms of
information about the program, including its syntactic structure, its type declarations and their
subtype / supertype relations, the types of its variables, and information about its data and
control flow patterns. By querying this database using a language superficially similar to SQL, the
system can be used to answer a variety of ad-hoc queries about the codebase.

Guided by the user's input, \Tool{} repeatedly queries the LLM until it synthesizes a CodeQL query
that is subsequently discharged over the entire codebase. The result of the query is presented to
the user in tabular form as the system output. The user may click on the presented program locations
to directly navigate to different parts of their codebase. The final CodeQL query is also included
as justification for the output presented.

We show the query generated in response to the ``\emph{constructors calling overridable methods}''
prompt in Figure~\ref{sfig:motiv:met05-j-ql}. In our user study, participants in the control group
(i.e., without access to \Tool) frequently expressed concern about the possibility of missed program
locations. On the other hand, even though none of the participants had prior experience with CodeQL,
a cursory inspection of the query provided them with confidence regarding the exhaustiveness of the
analysis results. Unlike purely LLM-based approaches, this queries therefore enable a repeatable,
auditable and exhaustive analysis of the entire codebase.

As discussed in the introduction, the primary challenge with realizing this approach is that CodeQL
is a highly expressive, low resource language with ``delicate'' semantics (i.e., even slight
variations can result in ill-formed queries or queries with substantially different semantics). In
the next section, we show how a combination of retrieval-augmented generation (RAG), compiler
feedback, self tests, and assistive queries can be used to overcome this difficulty.


\mclearpage \section{Workflow of the \Tool{} System}
\label{sec:alg}

\begin{figure*}
\includegraphics[width=\textwidth]{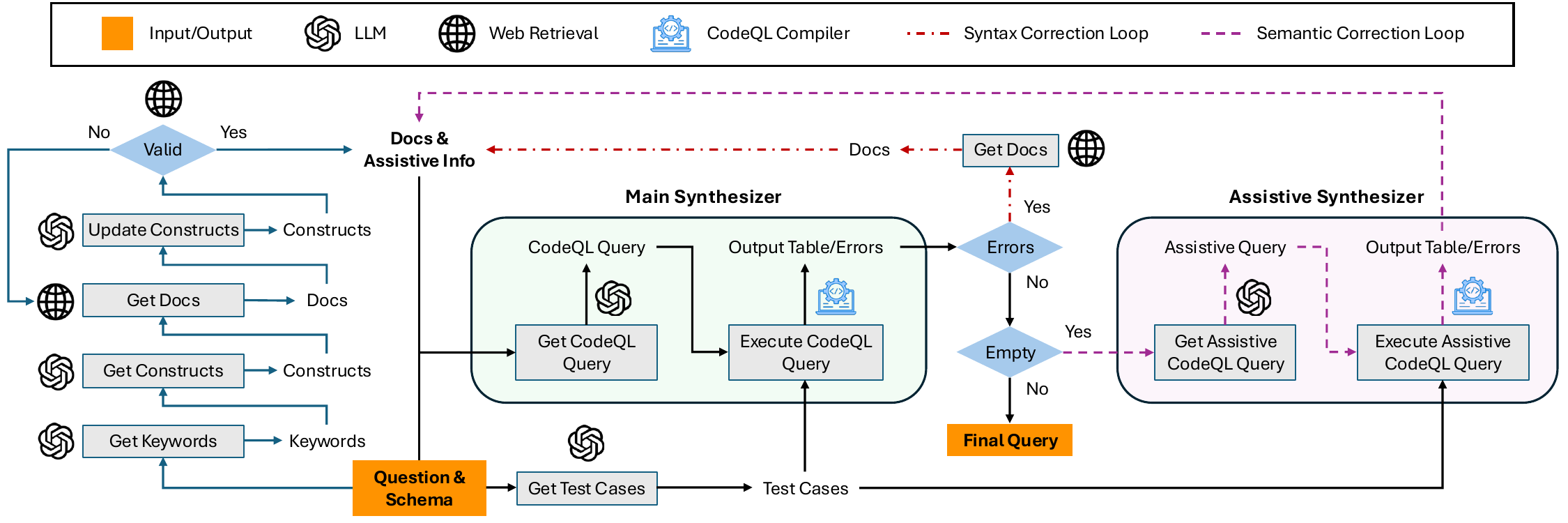}
\caption{
  Overall architecture of \Tool{}.
  \Tool{} first uses an LLM to retrieve relevant documentation and generate test cases that reflect
  the user's goal.
  It then repeatedly uses the LLM to generate a candidate CodeQL query and addresses syntax errors
  using RAG-based debugging and resolves semantic errors by issuing assistive queries.
  The final query is executed the entire codebase in order to produce the final table that is
  presented to the user.}
\label{fig:alg:arch}
\end{figure*}

We start by presenting the high-level workflow of \Tool{} in Figure~\ref{fig:alg:arch}.
\Tool{} receives a natural language goal, the schema of the desired output table, and a codebase as
input, and produces an output table as the result.
The algorithm consists of three main components: a preprocessing step, a retrieval-augmented
generation based query debugging phase, and a self-test based query verification mechanism. We now
describe each of these components in detail. All prompt templates may be found in the directory
named \texttt{prompts/merlin} in the attached artifact.

\mmclearpage
\subsection{Setup and Running Example}
\label{sub:alg:example}

As another example, say the programmer wishes to find all locations in their codebase where the
\lstinline|Object.equals()| method is used to compare two arrays. Although arrays in Java permit
comparison using the \lstinline|Object.equals()| method, this merely checks for reference equality,
rather than an elementwise comparison of their contents.
Programmers who wish to compare the \emph{contents} of two arrays must instead use the static
two-argument \lstinline|Arrays.equals()| method. This method considers two arrays equal if both
arrays contain the same sequence of elements, themselves compared according to
\lstinline|Object.equals()|.
To test for reference equality, it is recommended to use the reference equality operators,
\lstinline|==| and \lstinline|!=|. Because of the potential for confusion, the SEI CERT Coding
Standards therefore discourage the use of \lstinline|Object.equals()| to compare arrays.

The programmer writes the goal and the output table schema. Here, they may specify their goal as:
\guillemotleft{}\emph{Identify all method calls of the form \lstinline|array1.equals(array2)|, where
\lstinline|array1| and \lstinline|array2| are two arrays.}\guillemotright{}
They may specify the desired output schema as a simple one-column table:
\guillemotleft{}\emph{The location of the method call.}\guillemotright{}

\begin{figure*}
\centering
\begin{subfigure}[b]{.23\textwidth}
\centering
\begin{lstlisting}[style=fSQL, basicstyle=\tiny\ttfamily]
public class myClass {
 public static void main(String[] args) {
  Object[] arr1 = {1, 2, 3};
  int[] arr2 = {1, 2, 3};
  int[] arr3 = {4, 5, 6};
  boolean r1 = arr1.equals(arr2);
  boolean r2 = arr1.equals(arr3);
  boolean r3 = arr1.equals(arr1);
  System.out.println(r1);
  System.out.println(r2);
  System.out.println(r3);
 }
}
\end{lstlisting}
\caption{}
\label{sfig:alg:example:test}
\end{subfigure}
%
\hspace{.01\textwidth}
\begin{subfigure}[b]{.23\textwidth}
\centering
\begin{lstlisting}[style=fSQL, basicstyle=\tiny\ttfamily]
import java
from MethodCall c, Variable v1, Variable v2
where c.getMethod().getName() = "equals" and
c.getQualifier() = v1.getAnAccess() and
c.getAnArgument() = v2.getAnAccess() and
v1.getType().(ArrayType) and
v2.getType().(ArrayType)
select c.getLocation()
\end{lstlisting}
\caption{}
\label{sfig:alg:example:qwrong}
\end{subfigure}
%
\hspace{.01\textwidth}
\begin{subfigure}[b]{.23\textwidth}
\centering
\begin{subfigure}[b]{\textwidth}
\centering
\begin{lstlisting}[style=fSQL, basicstyle=\tiny\ttfamily]
import java
from Call c, Method m,
Expr e1, Expr e2
where c.getCallee() = m and
m.hasName("equals") and
c.getAnArgument() = e2 and
c.getQualifier() = e1 and
e1.getType().hasName("int[]") and
e2.getType().hasName("int[]")
select c.getLocation()
\end{lstlisting}
\caption{}
\label{sfig:alg:example:q1}
\end{subfigure}
\vspace{0.5ex}
\begin{subfigure}[b]{\textwidth}
\centering
\begin{lstlisting}[style=fSQL, basicstyle=\tiny\ttfamily]
import java
from MethodCall c
where c.getMethod().getName() = "equals"
select c.getQualifier().getType(),
c.getArgument(0).getType()
\end{lstlisting}
\caption{}
\label{sfig:alg:example:qassist}
\end{subfigure}
\end{subfigure}
\hspace{.01\textwidth}
\begin{subfigure}[b]{.2\textwidth}
\centering
\begin{lstlisting}[style=fSQL, basicstyle=\tiny\ttfamily]
import java
from MethodCall c, Expr e1,
Expr e2, Type t1, Type t2
where
c.getMethod().getName() = "equals" and
c.getQualifier() = e1 and
c.getArgument(0) = e2 and
e1.getType() = t1 and
e2.getType() = t2 and
t1 instanceof Array and
t2 instanceof Array
select c.getLocation()
\end{lstlisting}
\caption{}
\label{sfig:alg:example:finalq}
\end{subfigure}
%
\caption{
  Our running example in Section~\ref{sec:alg}.
  (\ref{sfig:alg:example:test}): The test case generated by the LLMs with an example use of
  \lstinline|Object.equals()| to compare arrays. The snippet contains three violations of the coding
  guideline, when each of the variables \lstinline|r1|, \lstinline|r2| and \lstinline|r3| are
  initialized.
  (\ref{sfig:alg:example:finalq}): The final CodeQL query generated by \Tool~which successfully
  detects all three violations of the guideline.
  (\ref{sfig:alg:example:qwrong}): The first CodeQL query in \Tool{}’s workflow, which contains syntactic errors.
  (\ref{sfig:alg:example:q1}): Initial syntactically correct suggestion. Note that because this
  query is looking specifically for comparisons between arrays of type \lstinline|int[]|, it does
  not detect any of the violations in the test case.
  (\ref{sfig:alg:example:qassist}): The assistive query generated by \Tool. It prints the types
  of \lstinline|o1| and \lstinline|o2| for every method invocation of the form
  \lstinline|o1.equals(o2)|, and helps the LLM to discover that the objects \lstinline|o1| and
  \lstinline|o2|, despite needing to be arrays, need not necessarily only contain integers.}
\label{fig:alg:example}
\end{figure*}


\mmclearpage
\subsection{Question Preprocessing}
\label{sub:alg:preproc}

\Tool{} begins with a preprocessing phase that obtains two key pieces of information:
\begin{inparaenum}[(\itshape a\upshape)]
\item a set of test cases that reflects the user's goal, and
\item a subset of the CodeQL documentation covering potentially relevant library constructs.
\end{inparaenum}
Both of these automatically obtained pieces of information will be used to guide the generation of
the analysis query by subsequent modules in the \Tool{} workflow.


\paragraph{Automatic generation of ``self-tests''}

An important failure mode for LLM-generated static analyzers is their tendency to produce empty
outputs. This is because, even when the query is syntactically well-formed, inherent ambiguities in
natural language and the delicate semantics of CodeQL queries lead to candidate queries with
degenerate behavior. We see an example in Figure~\ref{sfig:alg:example:q1} which---despite being
otherwise well-formed---fails to detect any of the violations in Figure~\ref{sfig:alg:example:test}.

We therefore start by prompting the language model to write test cases for a potential static
analyzer with the user-provided goal. The snippet in Figure~\ref{sfig:alg:example:test} is an
example of an automatically generated test case for the question regarding array equalities. We
subsequently use these automatically generated snippets as a simple litmus test that flags an
incorrect understanding of the goal by the ultimately synthesized program analyzer.

Note also that we use the same session for the entire interaction process with the LLM. As a result,
in addition to testing the analyzer, these automatically generated test cases also provide context
to improve the quality of LLM responses. As we will observe in our experimental evaluation, these
self-tests have the greatest impact on the effectiveness of \Tool.


\paragraph{Documentation retrieval}

Next, we hope to use a RAG-based approach~\cite{RAG} to provide the LLM with documentation about the
necessary CodeQL constructs during the query generation phase. However, we find that the language
model frequently hallucinates when asked to directly list / request the necessary documentation.
This is because focusing solely on concepts mentioned in the query leads to a listing of
non-existent / irrelevant constructs within the CodeQL standard library.

We instead follow an iterative refinement loop, where we first ask the LLM to identify keywords in
the user's question. For the case of our running example, the LLM identifies keywords such as
``\emph{identify}'', ``\emph{method}'', ``\emph{calls}'', ``\emph{array1}'', ``\emph{equals}'',
``\emph{array2}'', ``\emph{compare}'', and ``\emph{arrays}''.
Next, we ask the language model to identify constructs from the CodeQL standard library that
correspond to these keywords. This results in the list: \lstinline|MethodCall|,
\lstinline|MethodCall::getReceiver|, \lstinline|MethodCall::getArgument|, \lstinline|ArrayType|, and
\lstinline|ArrayType::getElementType|.
%
We validate each proposed construct by consulting the online CodeQL documentation. In the case of
our example, we discover that the class \lstinline|ArrayType| and the predicates
\lstinline|MethodCall::getReceiver| and \lstinline|ArrayType::getElementType| do not exist. We
therefore provide feedback about their non-existence to the LLM, and ask for an updated list of
constructs. We iterate until we obtain a list of complete and valid constructs from the LLM.

Another appealing aspect of this iterative documentation retrieval process is that it aligns more
closely with the hierarchical structure of the CodeQL language reference. For example, analyzer
classes such as \lstinline|MethodCall| and \lstinline|Array| are identified first, and their members
can be readily identified using documentation from the parent. For example, the documentation for
\lstinline|MethodCall| contains references to its predicates such as
\lstinline|MethodCall::getArgument|, thereby reducing the possibility of hallucination~\cite{codeql-java-methodcall}.


\mmclearpage
\subsection{RAG-Based Query Debugging}
\label{sub:alg:debugging}

Having identified the potentially useful constructs for synthesizing a program analyzer, we now
leverage these constructs to generate a well-formed query.
We start by prompting the LLM to generate an initial CodeQL query based on the user-specified goal,
the desired output table schema, the synthesized test cases, and the relevant retrieved
documentation.
This initial query is then compiled using the CodeQL compiler.
In most cases, the generated query contains one or more syntax errors and misuses of the standard
library. For instance, despite our previous discovery that the predicate \lstinline|ArrayType| does
not exist, the candidate query in Figure~\ref{sfig:alg:example:qwrong} hallucinates its existence and therefore fails to compile.

To address such errors, we extract the compiler's error messages and use them to guide a targeted
documentation retrieval from the CodeQL language reference. In this case, we fetch the documentation
for \lstinline|Expr::MethodCall| and use it to prompt the LLM to revise the query accordingly. If
the revised query compiles successfully, we move to the next stage. Otherwise, we iterate using the
new error messages to fetch more relevant documentation and prompt further revisions. We show the
final refined, successfully compiled query for this example in Figure~\ref{sfig:alg:example:q1}.


\mmclearpage
\subsection{Fixing Incorrect Analyzers with Assistive Queries}
\label{sub:alg:asq}

Unfortunately, as previously discussed in Section~\ref{sub:alg:preproc}, even syntactically
well-formed queries are overly selective and frequently result in empty outputs when applied to both
the generated self-tests and the user's codebase at large. Indeed, this is the case with the
candidate analysis query in Figure~\ref{sfig:alg:example:q1}, which incorrectly assumes that both
\lstinline|array1| and \lstinline|array2| are arrays of integers, while in reality they may contain
objects of other types. In particular, instead of checking whether \lstinline|arg1.getType()| and
\lstinline|arg2.getType()| have the textual name \lstinline|"int[]"|, the query should instead check
whether the types are instances of \lstinline|Array| type.

Notably, the \lstinline|Array| class from the CodeQL standard library was not included in the final
list of classes identified by the system at the end of the query preprocessing phase. At this point,
we suggest to the LLM that we can discharge a query that will help it diagnose and fix the problem,
and request it to propose such an assistive query. In response, the LLM proposes a query similar to
that shown in Figure~\ref{sfig:alg:example:qassist}.

The role of the assistive query is similar to \lstinline|print| statements in exploratory
programming~\cite{kery2017exploring}: These queries are usually not very selective (i.e., have
permissive \lstinline|where| clauses), and help the system in discovering entities and possible
relations between them. In the case of the \lstinline|Object.equals| example, the assistive query of
Figure~\ref{sfig:alg:example:qassist} causes the language model to discover the \lstinline|Array|
class which forms a crucial ingredient in the final analysis query in Figure~%
\ref{sfig:alg:example:finalq}.


After feeding back the results of the assistive query to the LLM, we repeat the compile-test-assist
loop as shown in Figure~\ref{fig:alg:arch} until the query produces a non-empty output on the
litmus test. At this point, we discharge the last synthesized query on the entire codebase and
present the results to the user as output.

From a particular viewpoint, \Tool{}'s use of a compile-test-assist loop, as illustrated in
Figure~\ref{fig:alg:arch}, is similar to an AI agent. \Tool{} can be viewed as an agent that
interacts iteratively with the CodeQL and Java compilers within this loop to synthesize a well
typed and semantically correct program analyzer that answers a user's question about the codebase.

\mclearpage \section{Experimental Evaluation}
\label{sec:eval}

Recall that \Tool{} produces its output in the form of a table, identifying locations in the
codebase that match the user's question. Our evaluation therefore focuses on the following research
questions:
\begin{enumerate}[label=\textbf{RQ\arabic*.}, ref=RQ\arabic*, leftmargin=*]
\item \label{enu:eval:recall} Does \Tool's output include known answer locations in the codebase?
\item \label{enu:eval:comprehensiveness} Does \Tool{} identify previously unknown answer locations?
\item \label{enu:eval:time} How heavily (in terms of prompt length) does \Tool{} use the LLM? 
\end{enumerate}


\paragraph{Benchmarks}

Bug-finding and vulnerability detection systems are a particularly rich source of analytical
questions of the kind answered by \Tool. E.g., those discussed in Sections~\ref{sec:motiv} and~%
\ref{sec:alg}.

Our first benchmark is based on the 142 detectors included as part of SpotBugs, an open source static
analyzer which flags coding issues within Java projects~\cite{SpotBugs}.

Examples include detecting the use of floating-point variables as loop counters,
suspiciously named methods such as \lstinline|Object.equal| instead of \lstinline|Object.equals|, and
instances of double-checked locking.

For each of these benchmarks, we reused the textual description available on the website, and
added the desired output schema by hand.

The output schema is used solely to standardize results and enable consistent comparison across
tools, and includes basic information about which source locations to report and how.

The second set of benchmarks was published by the GitHub Security Lab~\cite{GitHubSecurityLab}. It includes
CodeQL queries
designed to warn of various categories of issues in code written in a range of languages, including
Java, C++, C\# and JavaScript.
We excluded 33 queries that were no longer supported by CodeQL, and
17 queries that were longer than 1,000~characters.
This resulted in a set of 40~benchmarks.
We manually wrote natural language descriptions for each of these queries.

We also used the provided CodeQL query as the reference implementation and
attempted to find issues in the example codebases provided in the Security Lab repository.

Overall, our benchmark suite consists of 182 natural language questions and
reference analyzers, and 13 codebases consisting of 220--5,785 files each and
spanning 148,690--6,414,820 lines of code. The questions and schemas may be found in the files named
\lstinline|task.txt| in the attached artifact.


\paragraph{Baselines}

We compare \Tool{} to two alternative approaches to answering questions using LLMs:
\begin{inparaenum}[(\itshape a\upshape)]
\item Submitting the codebase and question text to an LLM and directly requesting an answer.
  We choose Gemini 3 Pro and Claude Code Sonnet 4.5 for this purpose.\footnote{File upload limits
  preclude the use of ChatGPT.}
\item Submitting the question text to an LLM and requesting a corresponding CodeQL query.
  We use \lstinline|gpt-4o|, Gemini 3 Pro, and Claude Code Sonnet 4.5.\sadra{last sentence is repetitive}
\end{inparaenum}
We refer to these baselines as Gemini/Question, Claude-Code/Question, GPT-4o/CQL, Gemini/CQL, and Claude-Code/CQL, respectively.
We also compare the performance of \Tool{} to three ablated versions:
\begin{inparaenum}[(\itshape a\upshape)]
\item directly requesting gpt-4o for a CodeQL query (GPT-4o/CQL),
\item including the documentation generated as part of the preprocessing step before requesting the
  query (GPT-4o+Docs), and
\item using the compiler to validate the candidate queries for well-formedness (GPT-4o+Docs+Compiler).
\end{inparaenum}


\begin{figure*}
\centering
\begin{subfigure}[c]{.24\textwidth}
\centering
\includegraphics[width=0.95\columnwidth]{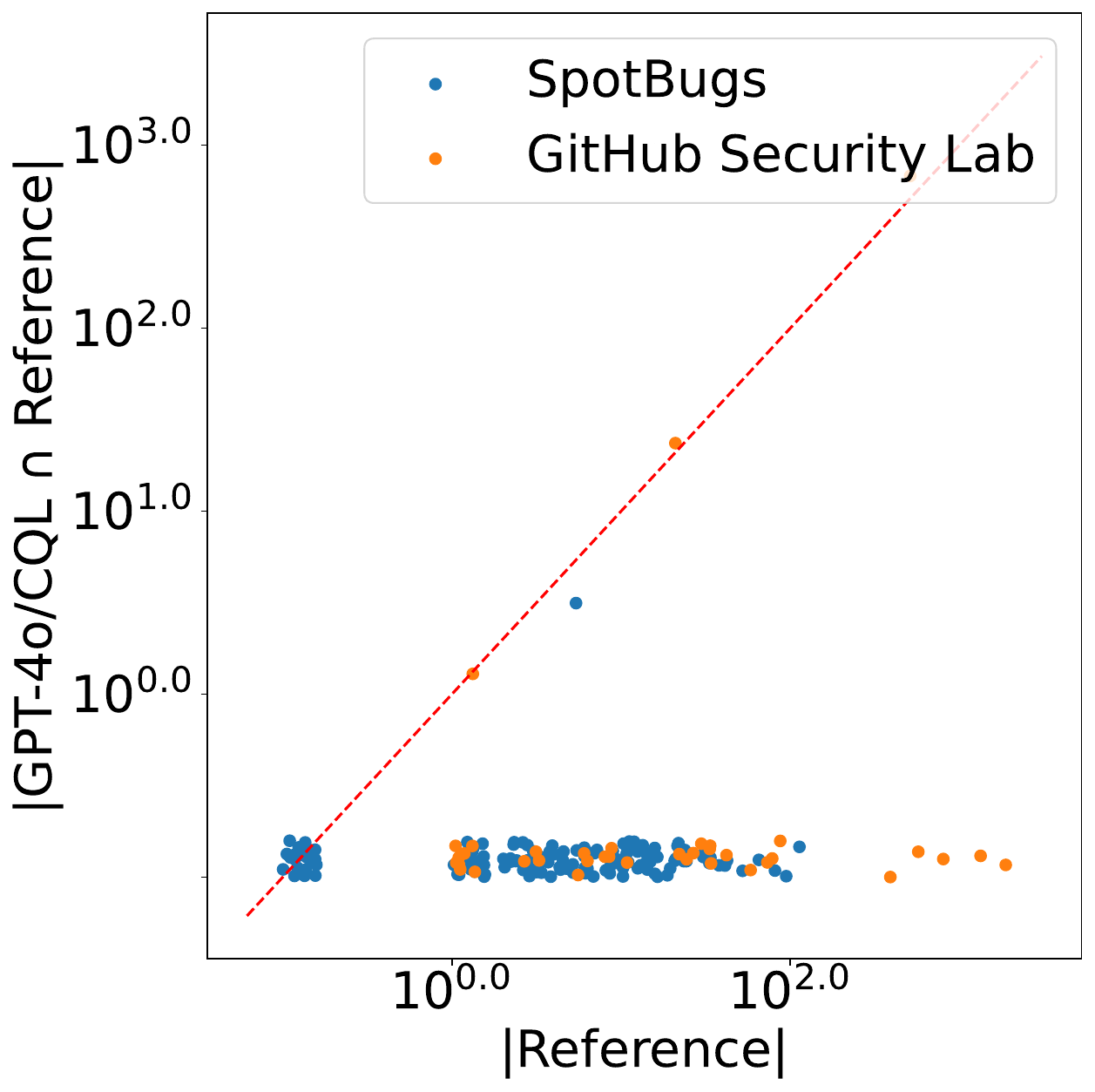}
\caption{}
\label{sfig:eval:plots1:alg-1-1}
\end{subfigure}
\hfill
\begin{subfigure}[c]{.24\textwidth}
\centering
\includegraphics[width=0.95\columnwidth]{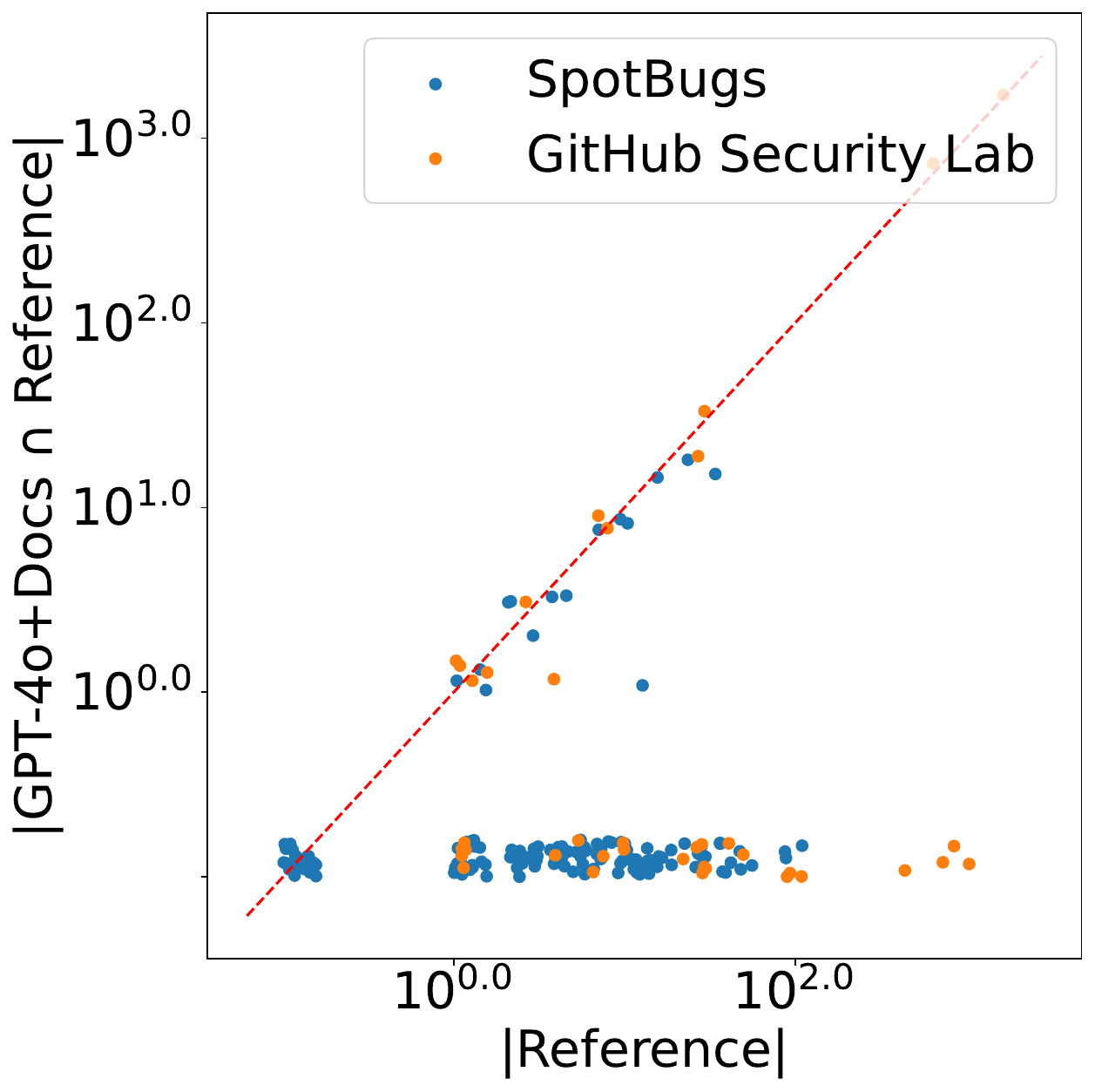}
\caption{}
\label{sfig:eval:plots1:alg-2-1}
\end{subfigure}
\hfill
\begin{subfigure}[c]{.24\textwidth}
\centering
\includegraphics[width=0.95\columnwidth]{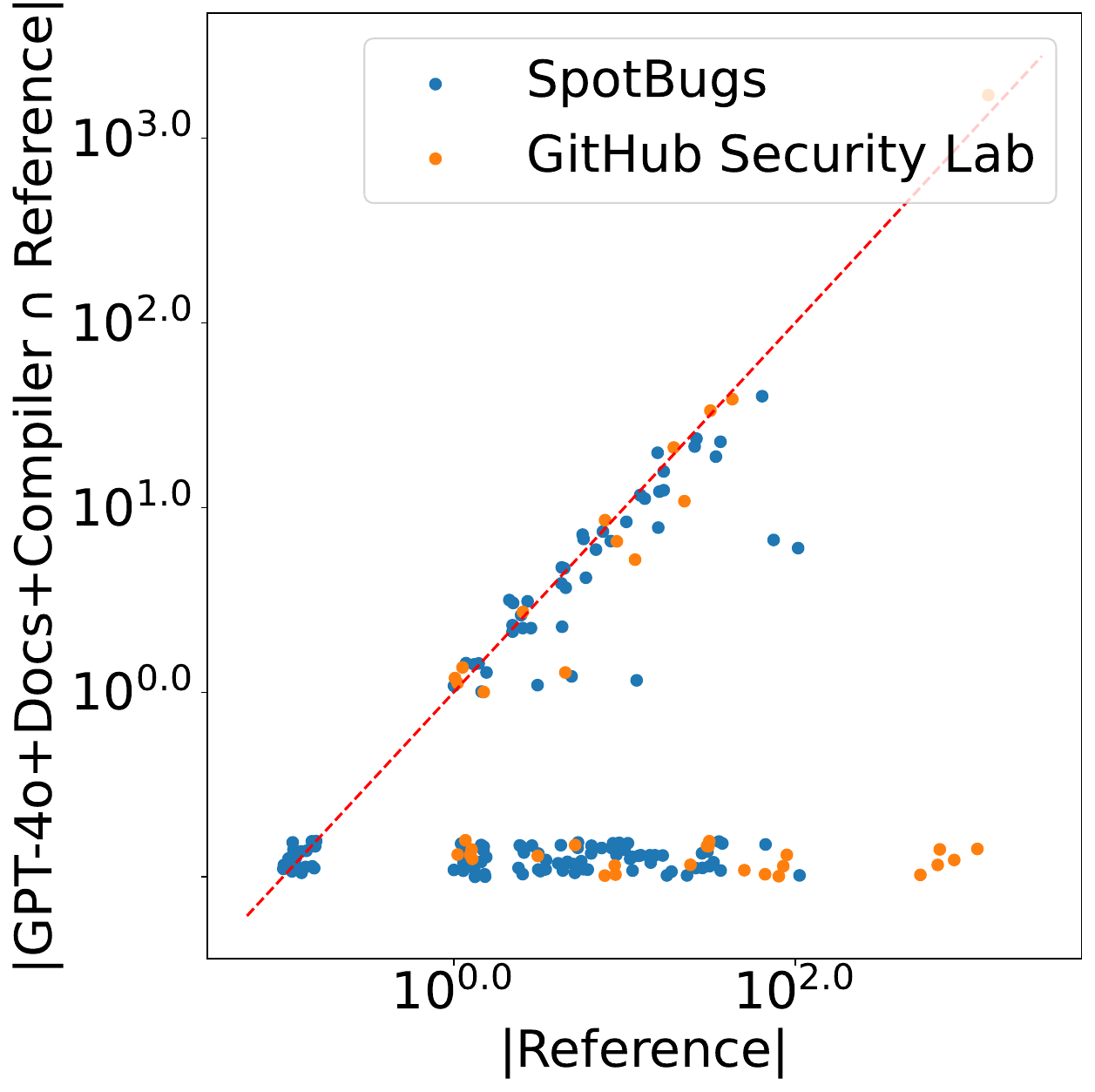}
\caption{}
\label{sfig:eval:plots1:alg-3-1}
\end{subfigure}
\hfill
\begin{subfigure}[c]{.24\textwidth}
\centering
\includegraphics[width=0.95\columnwidth]{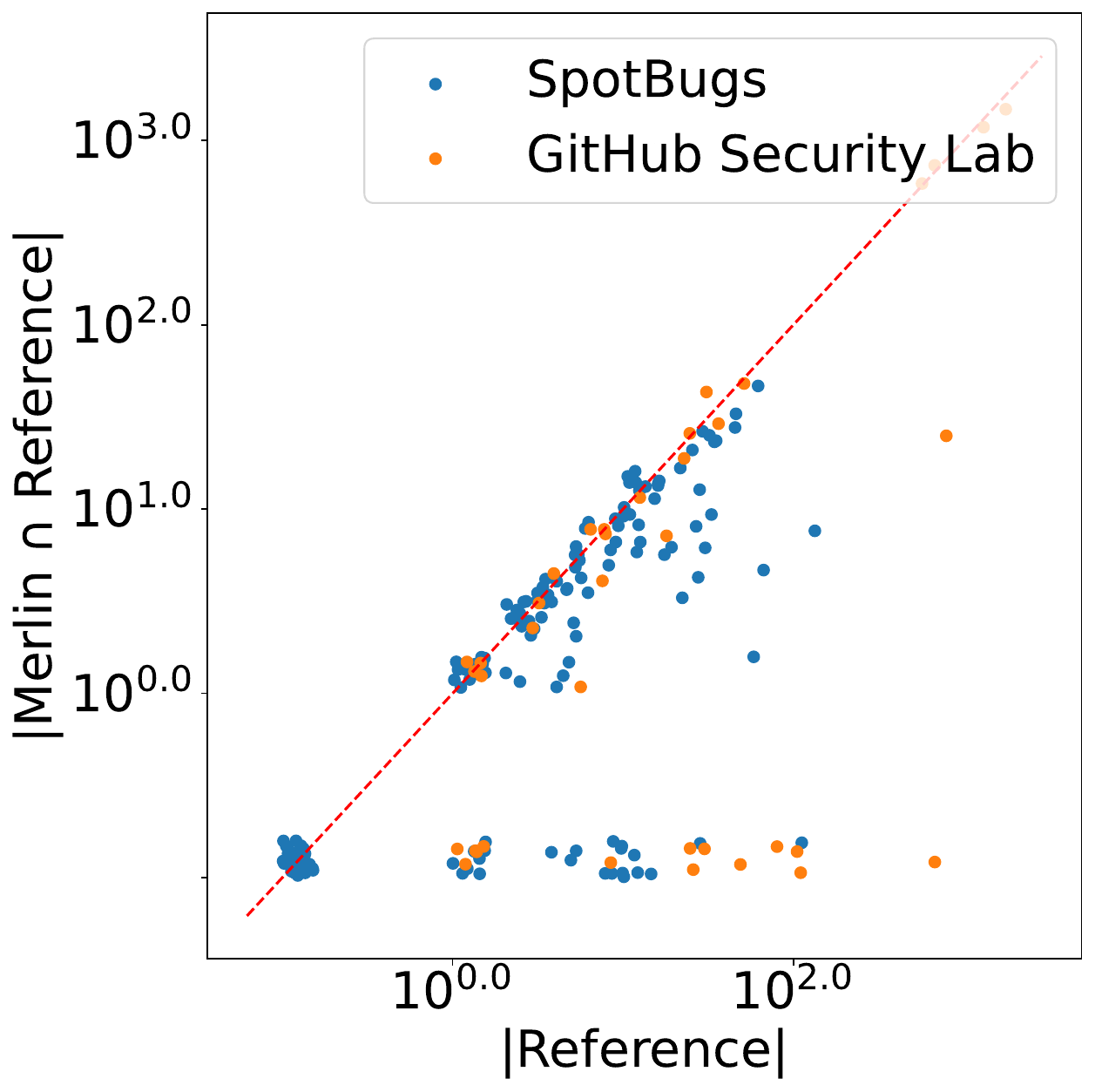}
\caption{}
\label{sfig:eval:plots1:merlin-1}
\end{subfigure}
\vspace{\baselineskip}
\begin{subfigure}[c]{.24\textwidth}
\centering
\includegraphics[width=0.95\columnwidth]{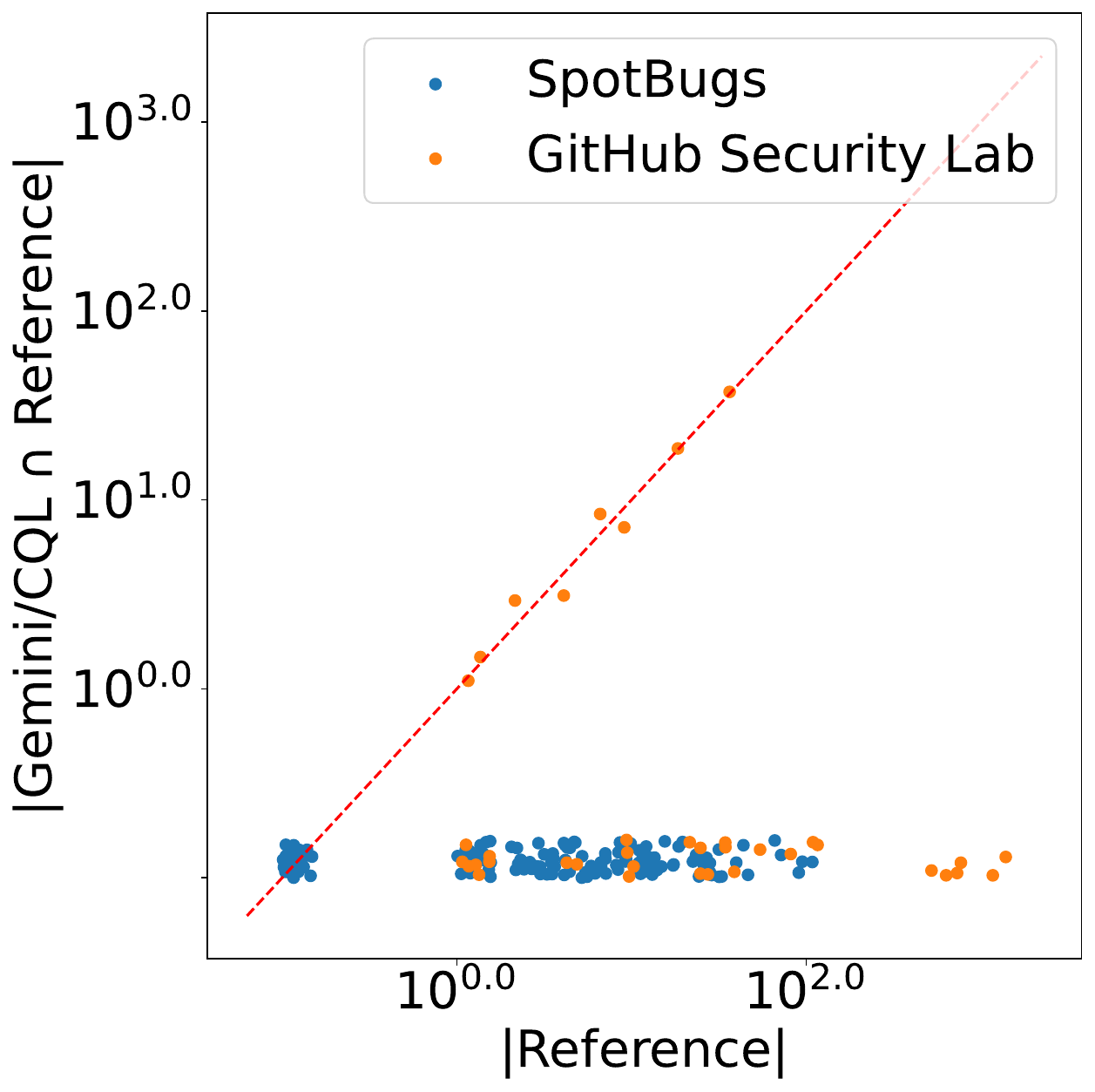}
\caption{}
\label{sfig:eval:plots1:gemini-query-1}
\end{subfigure}
\hfill
\begin{subfigure}[c]{.24\textwidth}
\centering
\includegraphics[width=0.95\columnwidth]{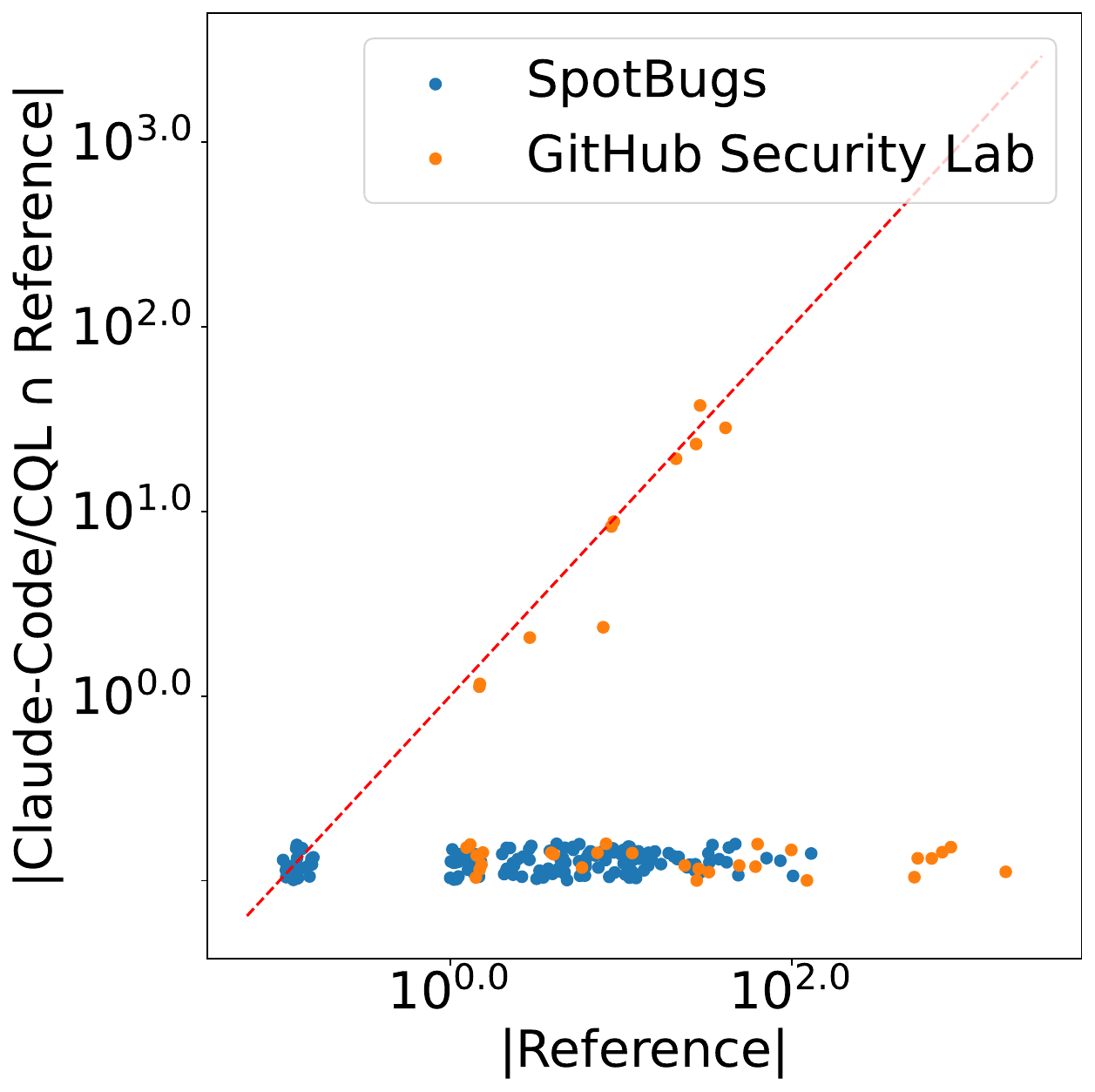}
\caption{}
\label{sfig:eval:plots1:claude-code-query-1}
\end{subfigure}
\hfill
\begin{subfigure}[c]{.24\textwidth}
\centering
\includegraphics[width=0.95\columnwidth]{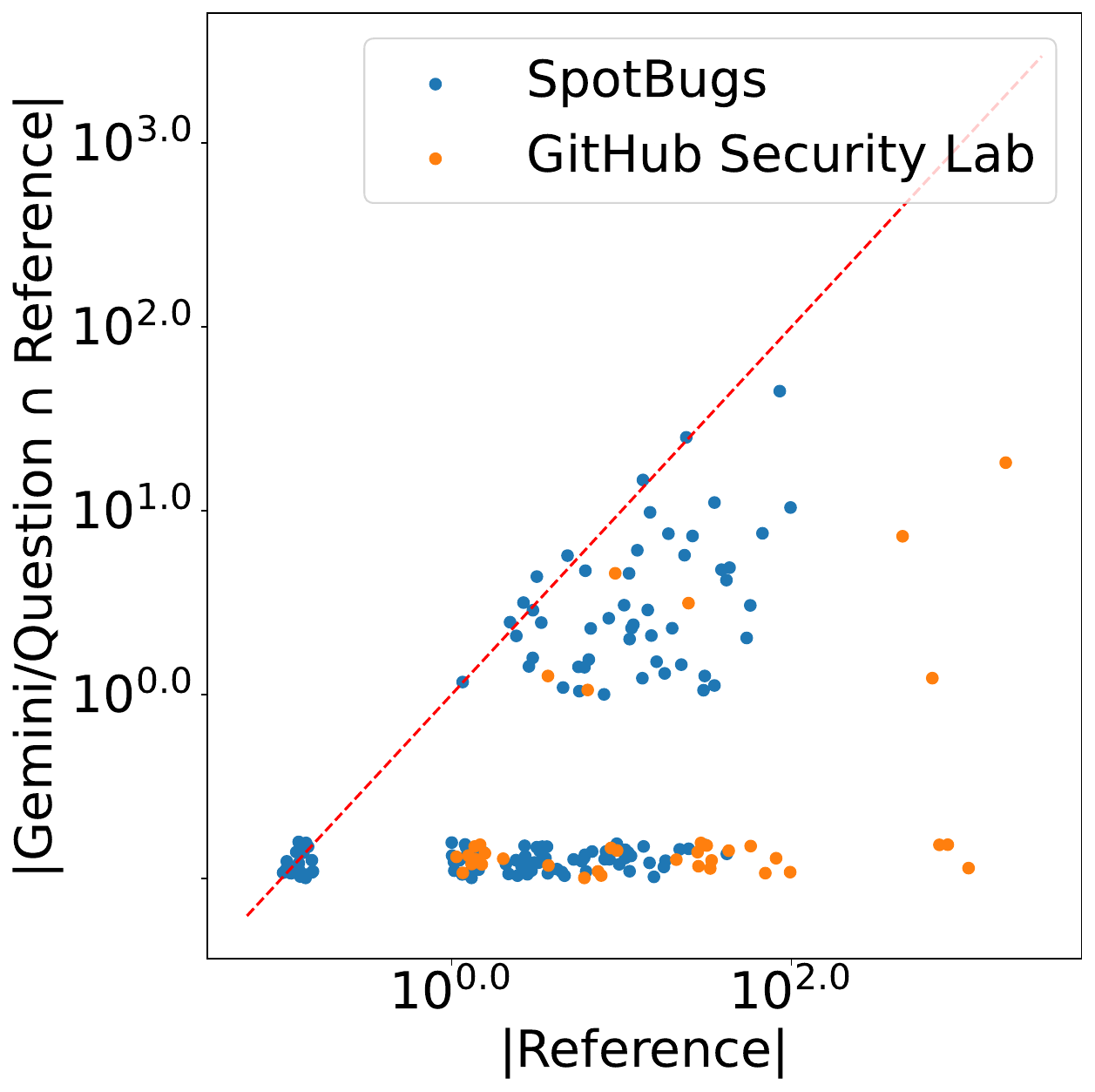}
\caption{}
\label{sfig:eval:plots1:gemini-question-1}
\end{subfigure}
\hfill
\begin{subfigure}[c]{.24\textwidth}
\centering
\includegraphics[width=0.95\columnwidth]{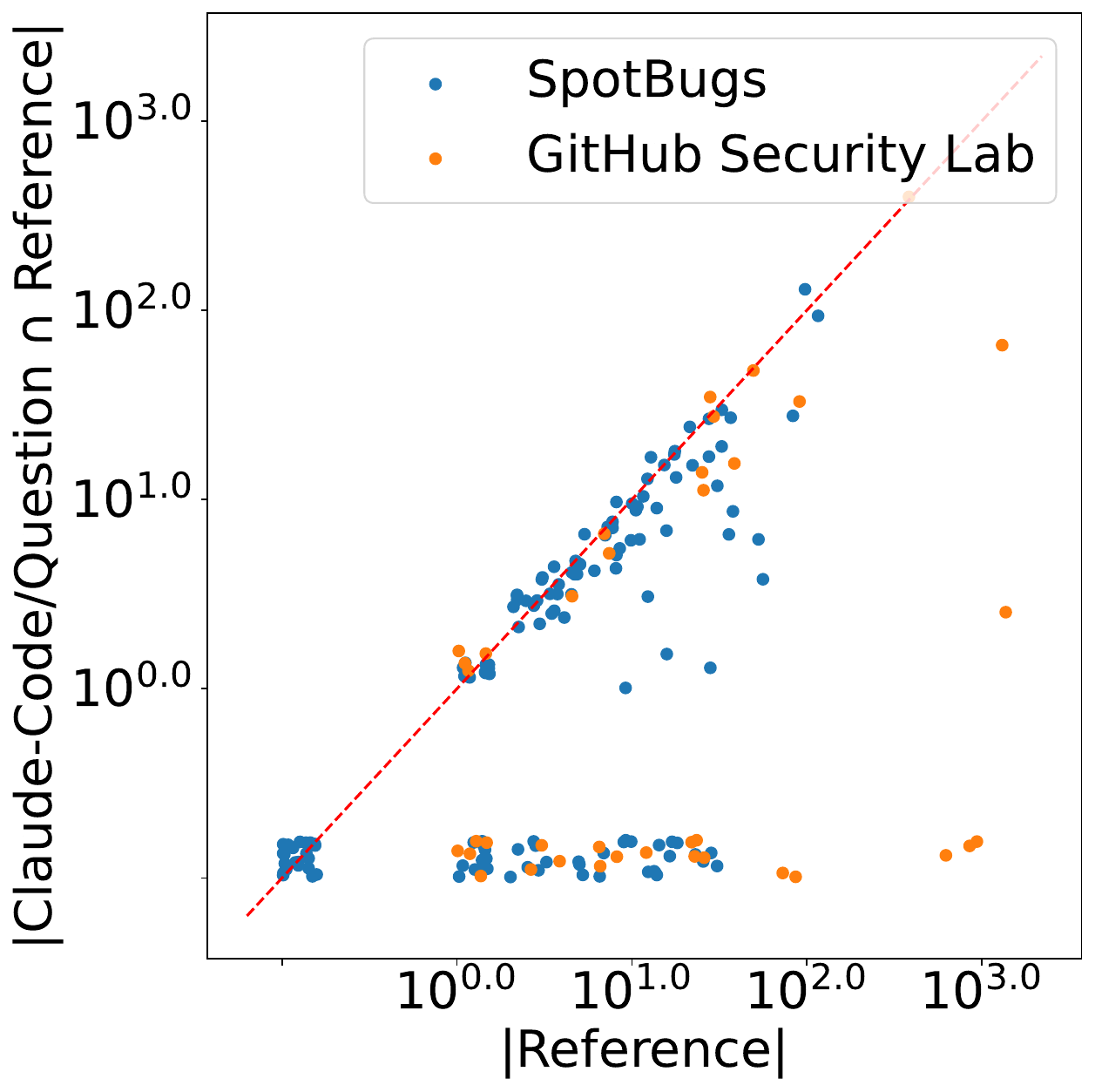}
\caption{}
\label{sfig:eval:plots1:claude-code-question-1}
\end{subfigure}
\caption{Overlap between the issues found by \Tool~(\ref{sfig:eval:plots1:merlin-1}),
  its ablated variants~(\ref{sfig:eval:plots1:alg-1-1}--\ref{sfig:eval:plots1:alg-3-1}),
  and the baselines~(\ref{sfig:eval:plots1:gemini-query-1}--\ref{sfig:eval:plots1:claude-code-question-1}),
  as compared to the reference solutions provided by SpotBugs and Security Lab, respectively.
  Each point indicates the number of locations identified by one benchmark/detector across the
  entire code repository.}
\label{fig:eval:plots1}
\end{figure*}

\begin{figure*}
\centering
\begin{subfigure}[c]{.24\textwidth}
\centering
\includegraphics[width=0.95\columnwidth]{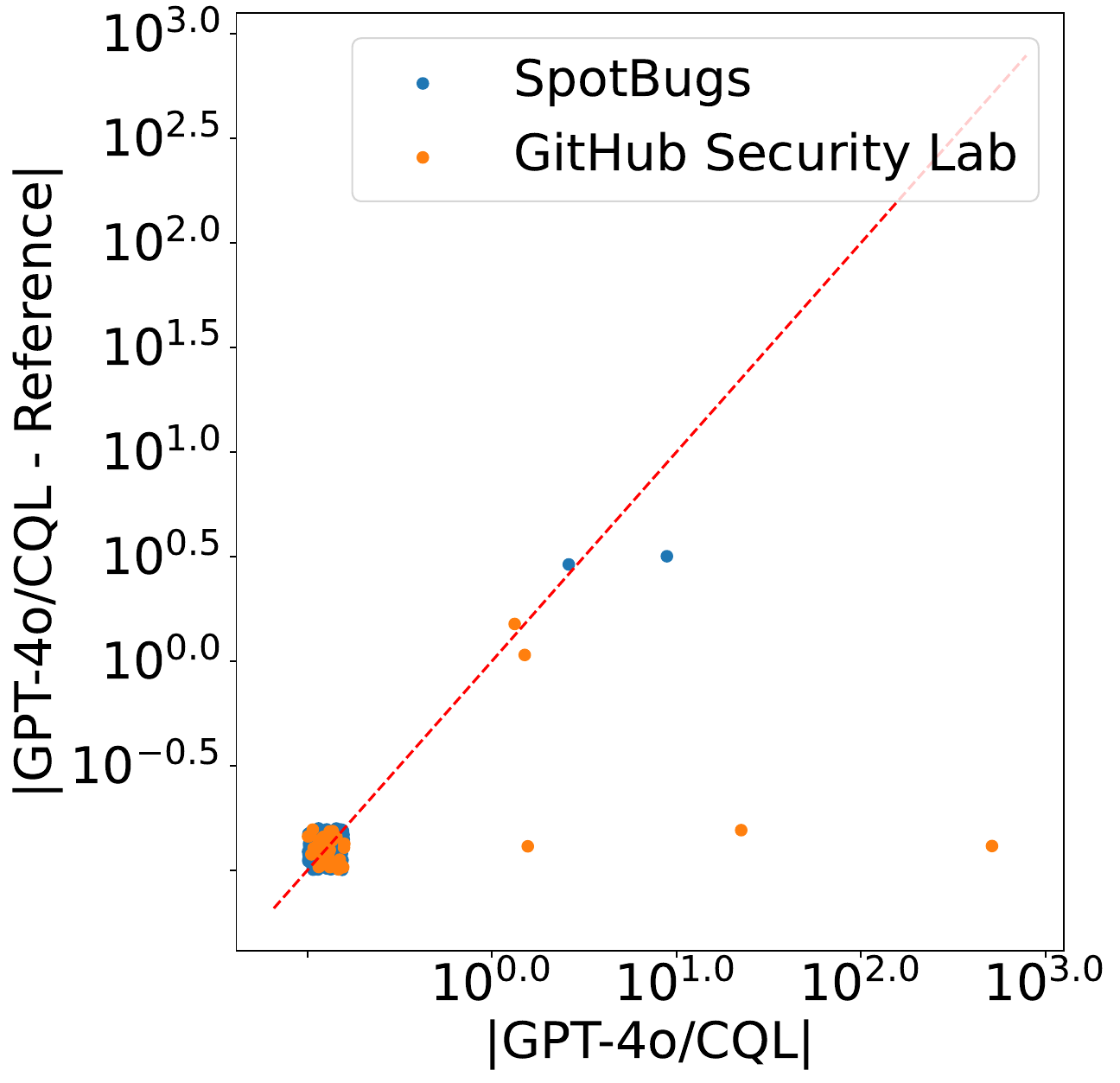}
\caption{}
\label{sfig:eval:plots2:alg-1-2}
\end{subfigure}
\hfill
\begin{subfigure}[c]{.24\textwidth}
\centering
\includegraphics[width=0.95\columnwidth]{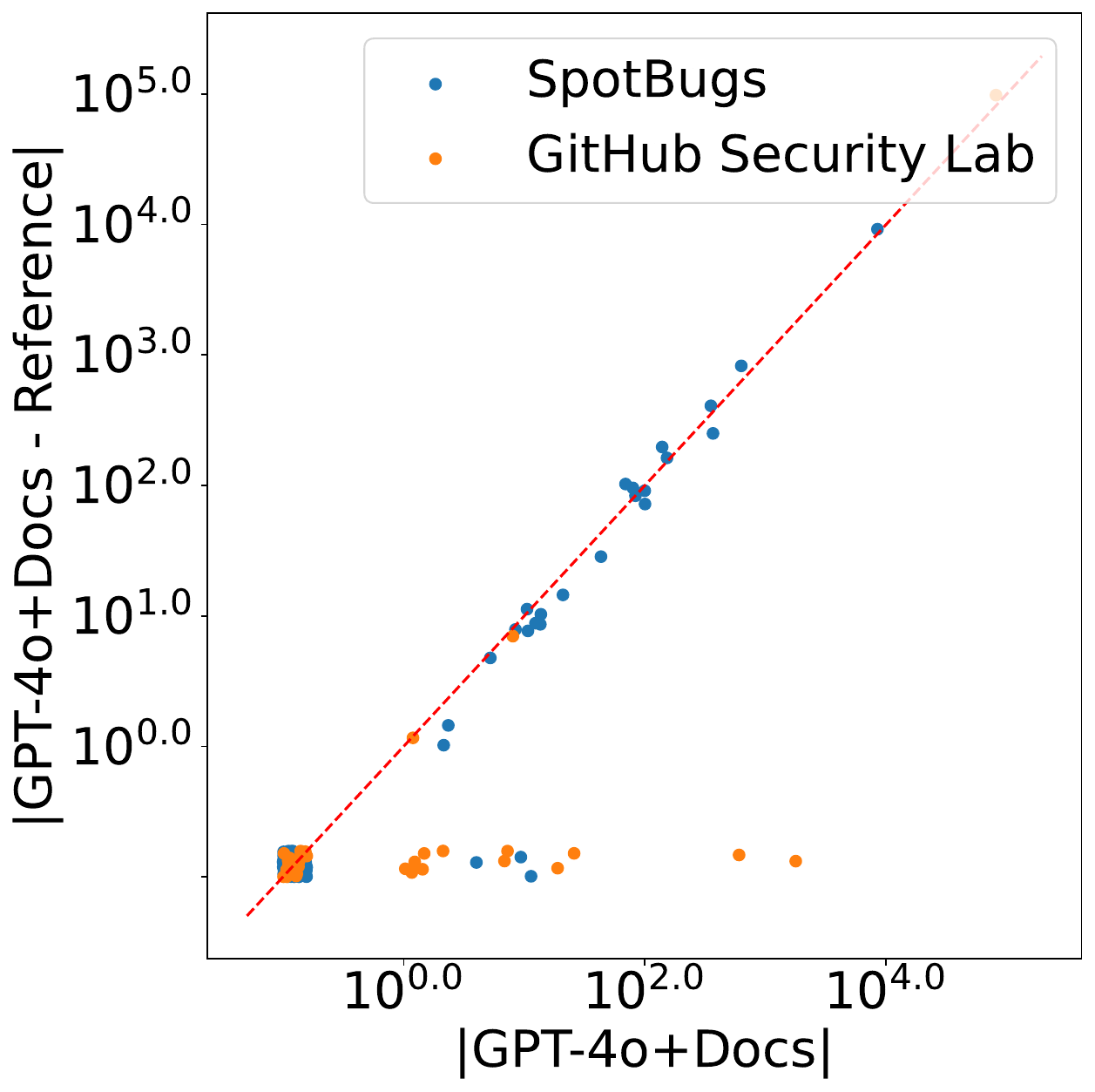}
\caption{}
\label{sfig:eval:plots2:alg-2-2}
\end{subfigure}
\hfill
\begin{subfigure}[c]{.24\textwidth}
\centering
\includegraphics[width=0.95\columnwidth]{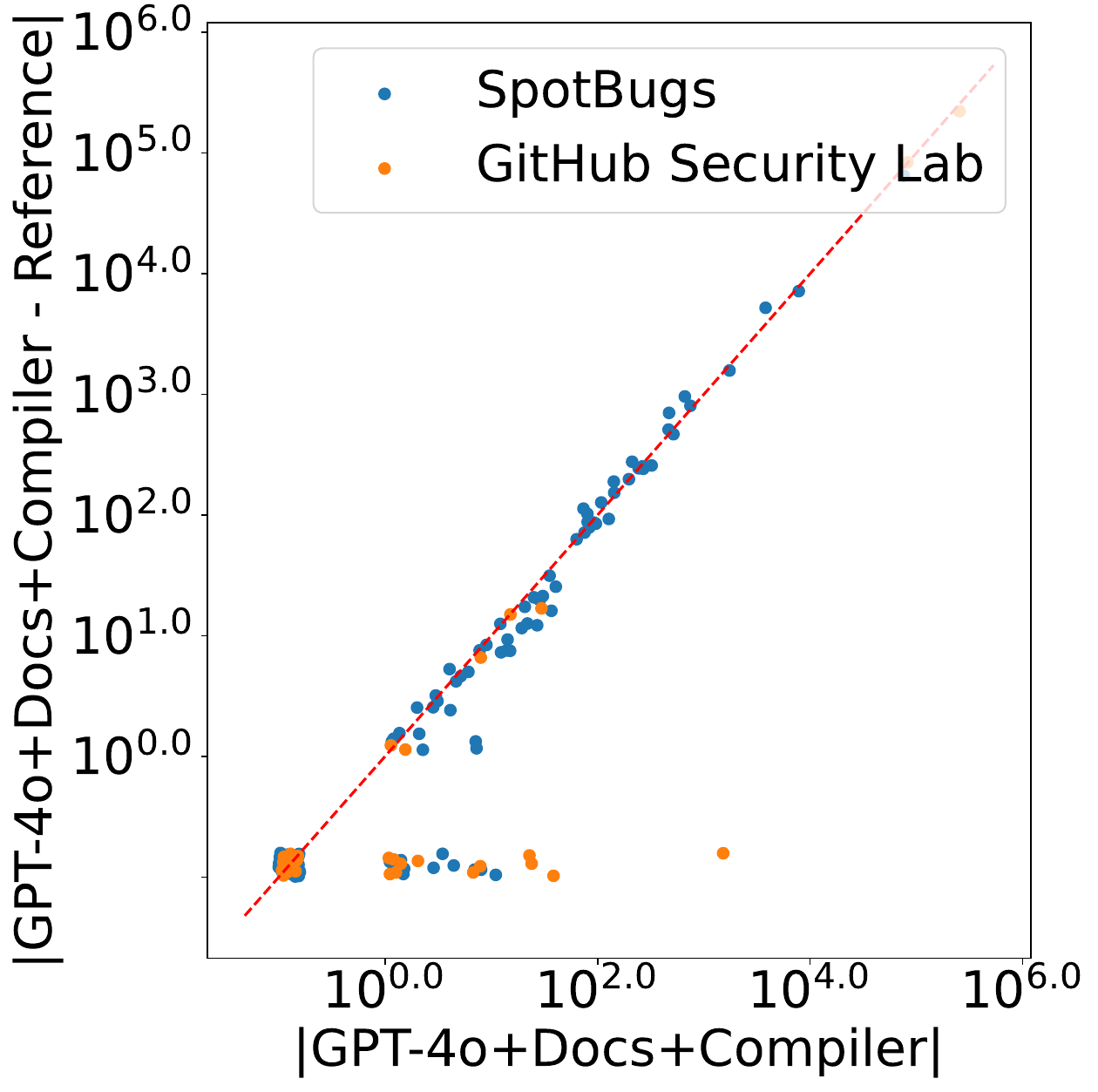}
\caption{}
\label{sfig:eval:plots2:alg-3-2}
\end{subfigure}
\hfill
\begin{subfigure}[c]{.24\textwidth}
\centering
\includegraphics[width=0.95\columnwidth]{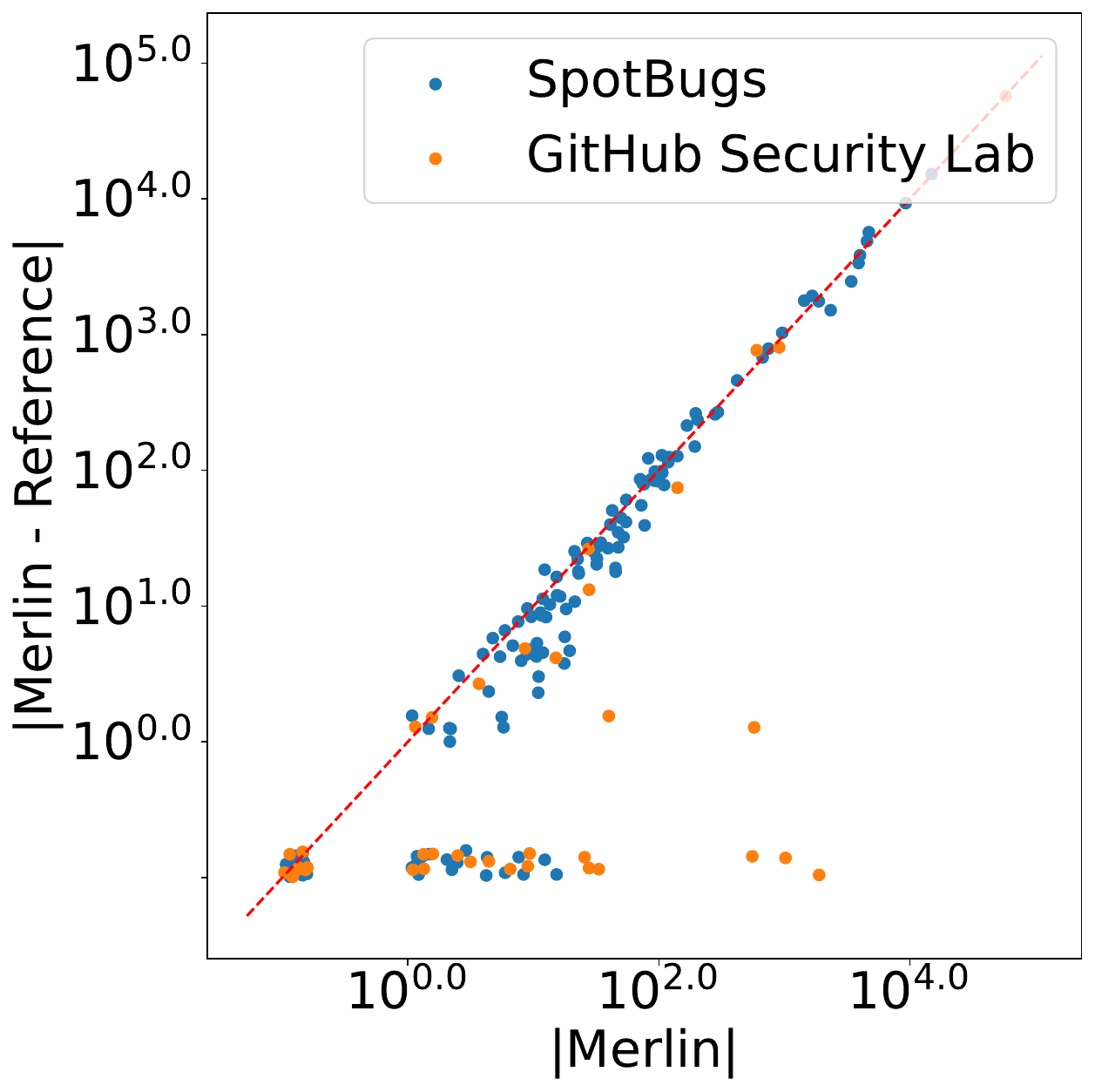}
\caption{}
\label{sfig:eval:plots2:merlin-2}
\end{subfigure}
\vspace{\baselineskip}
\begin{subfigure}[c]{.24\textwidth}
\centering
\includegraphics[width=0.95\columnwidth]{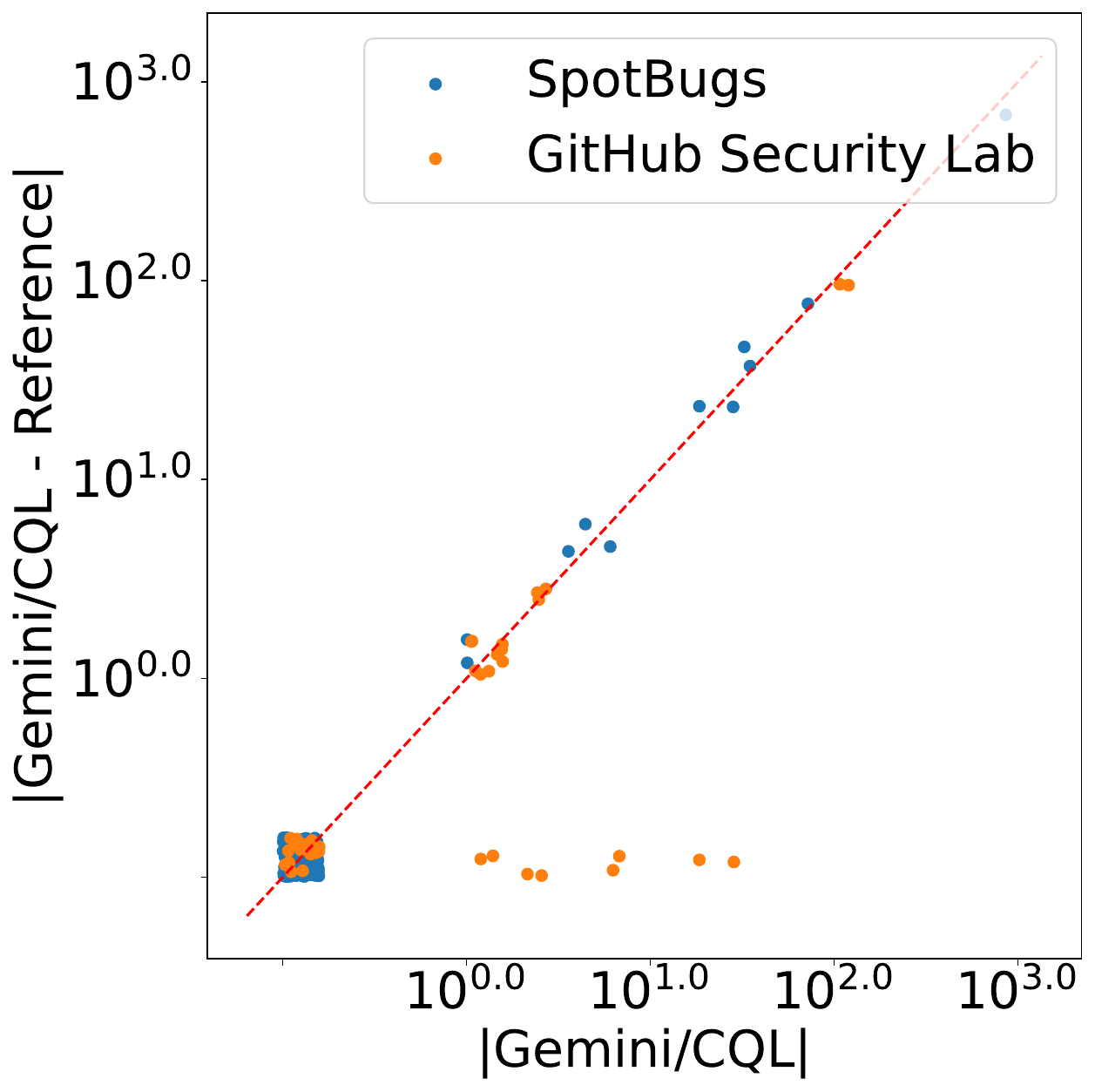}
\caption{}
\label{sfig:eval:plots2:gemini-query-2}
\end{subfigure}
\hfill
\begin{subfigure}[c]{.24\textwidth}
\centering
\includegraphics[width=0.95\columnwidth]{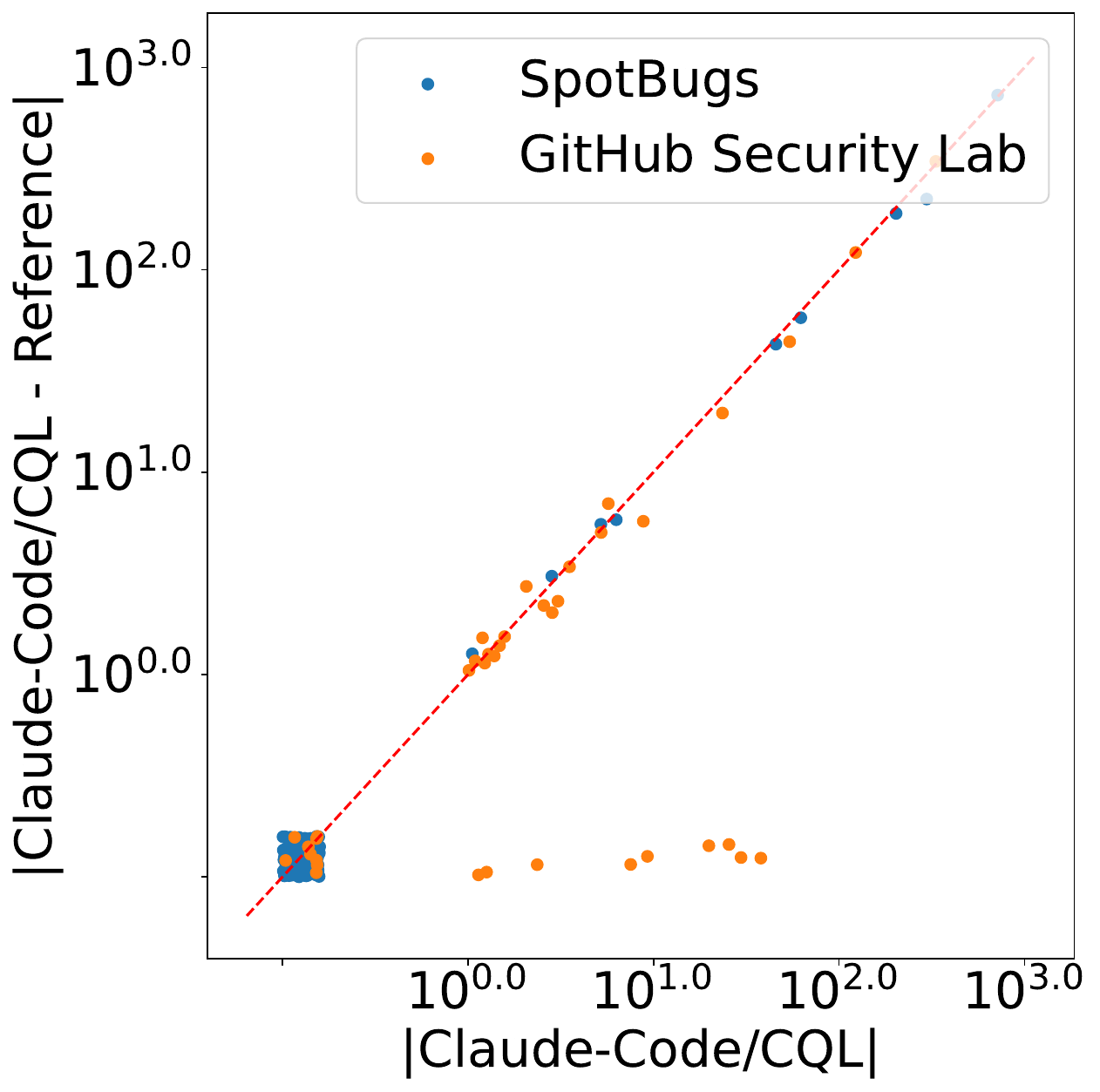}
\caption{}
\label{sfig:eval:plots2:claude-code-query-2}
\end{subfigure}
\hfill
\begin{subfigure}[c]{.24\textwidth}
\centering
\includegraphics[width=0.95\columnwidth]{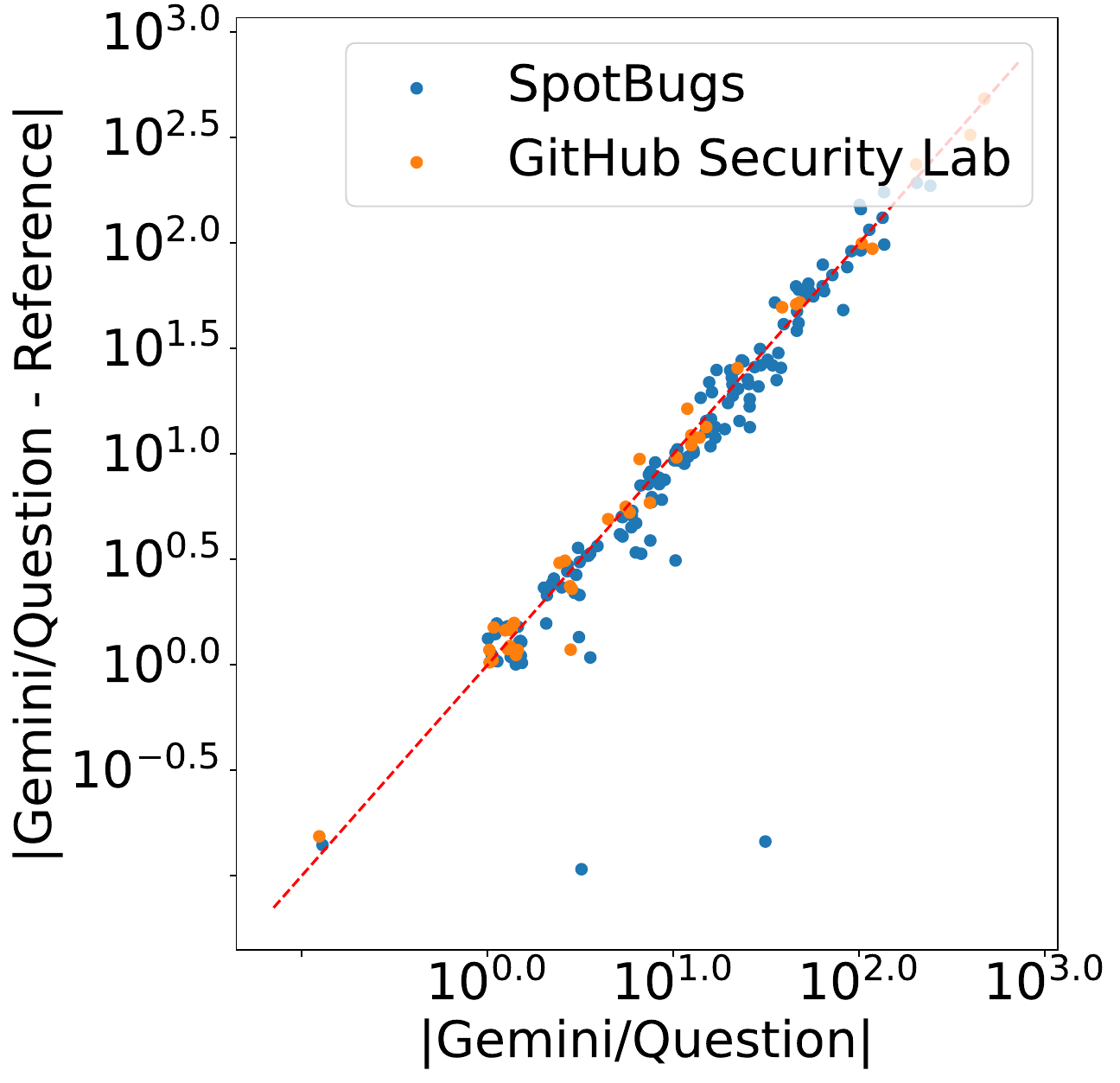}
\caption{}
\label{sfig:eval:plots2:gemini-question-2}
\end{subfigure}
\hfill
\begin{subfigure}[c]{.24\textwidth}
\centering
\includegraphics[width=0.95\columnwidth]{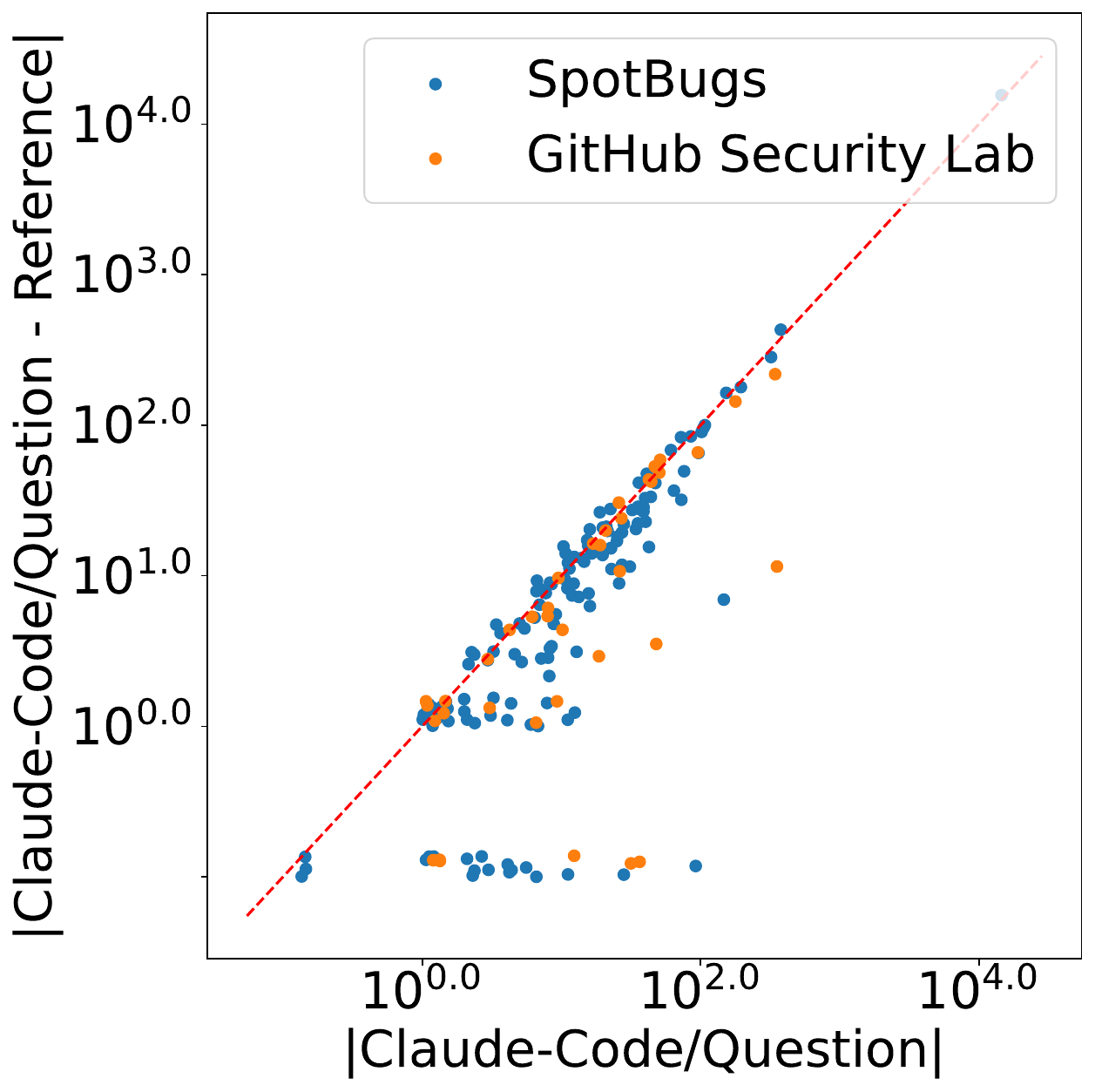}
\caption{}
\label{sfig:eval:plots2:claude-code-question-2}
\end{subfigure}

\caption{Proportion of new warnings (i.e., unreported by the reference SpotBugs and GitHub Security
  Lab analyzers) that are reported by
  \Tool~(\ref{sfig:eval:plots2:merlin-2}),
  its ablated variants (\ref{sfig:eval:plots2:alg-1-2}--\ref{sfig:eval:plots2:alg-3-2}), and
  the baselines~(\ref{sfig:eval:plots2:gemini-query-2}--\ref{sfig:eval:plots2:claude-code-question-2}) respectively.}
\label{fig:eval:plots2}
\end{figure*}

\begin{table}[t]
\caption{Macro- and micro- precision and recall statistics relative to the SpotBugs and Security Lab
  reference analyses and median total query length for each approach.
  The last column lists the total prompt length supplied to the LLM, aggregated as the median across
  all questions. While Claude Code does not provide fine-grained accounting, we estimate that each
  question costs 1.08\% and 0.04\% of the weekly usage limit for paid users, respectively. Numbers
  in parentheses indicate how many cases were removed due to division by 0.}
\label{tab:eval:statistics}
\centering
\small
\resizebox{\linewidth}{!}{
\begin{tabular}{lccccc}
\toprule
  \textbf{Algorithm} &
  \multicolumn{2}{c}{\textbf{Macro}} &
  \multicolumn{2}{c}{\textbf{Micro}} &
  \textbf{Size} \tabularnewline
  \cmidrule(lr){2-3} \cmidrule(lr){4-5}
  & \textbf{Prec.} & \textbf{Recall} & \textbf{Prec.} & \textbf{Recall} & \tabularnewline
\midrule
  Gemini/Question & 0.07 (2) & 0.13 (24) & 0.05 & 0.04 & 7.7 MB \tabularnewline
  Gemini/CQL & 0.24 (148) & 0.05 (24) & 0.06 & 0.01 & 451 bytes \tabularnewline
  Claude-Code/Question & 0.32 (3) & 0.50 (24) & 0.08 & 0.21 & - \tabularnewline 
  Claude-Code/CQL & 0.24 (143) & 0.06 (24) & 0.06 & 0.02 & - \tabularnewline 
\midrule
  GPT-4o/CQL & 0.50 (175) & 0.02 (24) & 0.99 & 0.08 & 451 bytes \tabularnewline
  GPT-4o+Docs & 0.44 (142) & 0.16 (24) & 0.02 & 0.32 & 22 KB \tabularnewline
  GPT-4o+Docs+Compiler & 0.35 (92) & 0.31 (24) & < 0.01 & 0.3 & 25 KB \tabularnewline
\midrule
  \Tool & 0.40 (32) & 0.62 (24) & 0.03 & 0.64 & 29 KB \tabularnewline
\bottomrule
\end{tabular}}
\end{table}


\mmclearpage \subsection{\ref{enu:eval:recall}: Does \Tool's Output Include Known Answer Locations in the Codebase?}
\label{sub:eval:recall}

For each benchmark question, we measured the overlap between locations identified by \Tool{} to those
flagged by the reference analyzers. We visualize our measurements in Figure~%
\ref{fig:eval:plots1}. In each figure, the $x$-axis represents the number of issues found
by the reference analyzer, while the $y$-axis represents the number of issues found by both the
reference analyzer and the target algorithm.

Points located near the $x = y$ line indicate that the target algorithm identifies most of the
issues found by the reference analyzer. From Figure~\ref{sfig:eval:plots1:alg-1-1} to
Figure~\ref{sfig:eval:plots1:merlin-1}, we observe a gradual increase in the number of points
near this line, indicating improved performance from each component of our approach. In particular,
Figure~\ref{sfig:eval:plots1:merlin-1} shows the largest cluster of points around the $x = y$ line,
suggesting that \Tool{} is effective at reproducing the results of the reference analyzers.

Quantifying the performance that we visually see in Figure~\ref{fig:eval:plots1} with concrete
recall values is tricky because
SpotBugs itself does not flag any warnings across the entire codebase for 24 of our benchmarks.
Naively calculating task-wise recall can therefore lead to divide-by-zero errors.
In addition, notice that the number of reported locations ranges across multiple orders of
magnitude across our 182~benchmark questions.
Aggregating the results and calculating a single recall value can lead to over-representation from a
subset of the benchmarks.
We therefore separately calculate macro- and micro-recalls of the different approaches~%
\cite{opitz-2024-closer}, which we report in Table~\ref{tab:eval:statistics}.

Notice that both macro- and micro-recall statistics improve significantly across the ablated
versions of \Tool{} from 0.02 to 0.62 and from 0.08 to 0.64 respectively. In addition, notice that
the most significant improvement comes from a combination of self-tests and assistive queries.

In contrast, Figures~\ref{sfig:eval:plots1:gemini-query-1} and~%
\ref{sfig:eval:plots1:claude-code-query-1} show that Gemini and Claude Code struggle to generate
CodeQL queries for answering these questions: they often fail to produce even syntactically
correct queries.

When asked to directly answer the question, notice that Gemini rarely detects issues identified by
the baseline (Figure~\ref{sfig:eval:plots1:gemini-question-1}), a result which is consistent with
its low macro- and micro-recall values. Anecdotally, many of its reported locations are meaningless
(e.g., closing braces around code blocks) or even point to empty lines in the codebase.

Claude Code performs better than other baselines in the direct question answering mode
(Figure~\ref{sfig:eval:plots1:claude-code-question-1}): This is because Claude Code is
an agentic LLM which incorporates the output of external tools such as \texttt{ls} and \texttt{grep}
as part of its reasoning process.
Still, it is less effective than \Tool{} at reproducing the results of the reference
analyzers (macro-recall: 0.5 vs. 0.62, micro-recall: 0.21 vs. 0.64).

A deeper investigation reveals that Claude Code is more effective in identifying locations from the
SpotBugs benchmarks (macro-recall: 0.55) than over the Security Lab dataset (macro-recall: 0.33).
While it is hard to say with certainty, we suspect that this is because the SpotBugs test suite
contains numerous hints about target locations, including conspicuously named files and variables
and helpful comments surrounding the code in question.

Finally, we investigated the points located far away from the $x = y$ diagonal in Figure~%
\ref{sfig:eval:plots1:merlin-1}. Many of these points are attributable to ambiguities in the
natural language questions rather than to failures of the \Tool{} query generation system itself.
One example is the benchmark description from the SpotBugs dataset:%
\footnote{\url{https://spotbugs.readthedocs.io/en/latest/bugDescriptions.html\#hrs-http-response-splitting-vulnerability-hrs-request-parameter-to-http-header}}
\guillemotleft{}\emph{Find code that constructs an HTTP Cookie using an untrusted HTTP parameter. If
  this cookie is added to an HTTP response, it will allow an HTTP response splitting
  vulnerability.}\guillemotright{}
Terms such as ``\emph{untrusted}'' and ``\emph{suspicious}'' are inherently vague and open to
multiple interpretations, which can lead to variations in how code patterns are identified.

\mmclearpage 
\subsection{\ref{enu:eval:comprehensiveness}: Does \Tool{} Identify Previously Unknown Answer Locations?}
\label{sub:eval:comprehensiveness}

Next, Figure~\ref{fig:eval:plots2} shows the fraction of reported locations that were not also reported by the reference
analyzers. In each figure,
the $x$-axis represents the number of locations reported by the evaluated algorithm, and the
$y$-axis shows the number of locations that were reported by the evaluated algorithm but not by
SpotBugs or the Security Lab queries.

These measurements may be quantified as the precision of the respective algorithms which we report
in Table~\ref{tab:eval:statistics}.

Figure~\ref{sfig:eval:plots2:merlin-2} shows that \Tool{} reports a number of previously
unreported locations. While it is tempting to classify these additional locations as being
spurious reports, a deeper investigation reveals that these frequently represent bugs that were
mistakenly not reported by the reference analyzer. For example, the following detector from SpotBugs:
\guillemotleft{}%
\emph{Find classes which are declared to be \lstinline|final|, but which also declare fields that
are \lstinline|protected|. Since the class is \lstinline|final|, it cannot be derived from, and the
use of \lstinline|protected| is confusing.}%
\guillemotright{}
fails to report obvious instances of this pattern,
including the following (from the file \texttt{ConfusingParenting.java} in the SpotBugs test
suite):
\begin{lstlisting}
public final class ConfusingParenting {
  protected int a;
  protected Object b;
}
\end{lstlisting}
This and numerous other similar observations suggested that there were discrepancies between the
documentation and what the actual SpotBugs analyzer was doing, and prompted us to conduct the survey in which we asked users to label the ground truth for a randomly chosen
subset of reported locations. The results suggest that many of the locations found by \Tool{} are in
fact relevant to the question, and that the precision measurements in Table~\ref{tab:eval:statistics} are likely to be
undercounts with respect to the actual ground truth.

\mmclearpage \subsection{\ref{enu:eval:time}: How Heavily Does \Tool{} Use the LLM While Answering Questions?}
\label{sub:eval:time}

Finally, we measured how extensively \Tool{} and each of the baseline approaches used the LLM.

The lowest LLM prompt lengths are observed for Gemini, Claude Code, and gpt-4o when they are asked
to generate CodeQL queries. This is expected, since the codebase is not provided as part of the
context in this setting, resulting in lower prompt lengths and a lighter processing workload. In
contrast, the highest context sizes occur for Gemini and Claude Code when they are asked to directly
answer questions. In this setting, both the codebase and the question are provided to the LLM, which
significantly increases the prompt lengths and computational burden.

This direct-question setting also introduced several practical challenges for Gemini and Claude
Code. Gemini enforces a limit of 1,000 files per upload, requiring us to split large codebases into
smaller sub-codebases. This process is time-consuming and error-prone, as dependent files may be
separated across different sub-codebases. Moreover, we had to run Gemini multiple times for each
sub-codebase, and in some cases the model terminated without producing a response.

Claude Code also posed substantial difficulties. It enforces rate limits both across 5-hour windows
and across each week. For our codebases, we were only able to ask only approximately 10~questions
per hour. Due to the weekly limits, completing the experiments required nearly two full weekly usage
cycles. In addition, Claude Code frequently requested permission to execute system commands, which
we had to manually review and individually approve due to privacy concerns.\sadra{this para is mostly tech challenges that may be solved by changing access level. I'd say we should remove it.}

In contrast, \Tool{} synthesizes program analyzers independently of the target codebase and
therefore does not require the codebase to be included in the context. As a result, \Tool{} exhibits
substantially lower context usage. Another notable advantage of \Tool{} over the other baselines is
that once a CodeQL query is synthesized, it can be reused across arbitrary codebases. In contrast,
the LLM-based baselines must be rerun for each new codebase, which is significantly more expensive.

\mclearpage \section{User Study on \Tool{}'s Usefulness}
\label{sec:user-study}

Finally, we designed a user study to validate the usefulness of \Tool{} in real-world software
development and compare it to
traditional LLMs and program analysis tools. We asked users to perform three tasks in which we presented them with a coding issue
and asked them to:
find instances of the issue in the given codebase, and
fix the code so as to remove the issue.
(Although the scope of \Tool{} is technically limited to only the ``\emph{answer questions}'' /
``\emph{find locations}'' phase, we included the ``\emph{fix code}'' directive simply because we
were curious to understand the impact of our system.)
We used this study to investigate the following research questions:
\begin{enumerate}[label=\textbf{RQ\arabic*.}, ref=RQ\arabic*, leftmargin=*, resume]
\item \label{enu:user-study:impact-process}
  How does \Tool{} impact the programmer's question answering process as compared to conventional
  program analysis or chat-based LLM tools?
\item \label{enu:user-study:effectiveness}
  How effective is \Tool{} in helping developers locate and fix code issues compared to other
  assistive techniques?
\end{enumerate}


\paragraph{Participants}

Once again, we recruited 18 participants through professional forums and the university mailing
list.
They ranged in age from 18 to 34 years.
17 were male and one female.
Participants included 13 students, 1 researcher, 3 professional software developers, and 1
participant with an occupation outside these categories.
11 participants had 6-10 years of programming experience.
All our participants had used IDE tools more than once.
Six participants were familiar with debuggers like GDB and LLDB, and some code analyzers like Valgrind.

\paragraph{Study tasks}

From the SEI CERT Oracle Coding Standard for Java~\cite{CertJava}, we shortlisted 8 coding
guidelines which each had fewer than 10 violations in the codebase, and which we determined were
feasible to identify and fix within the time allotted to each task in the user study (40 minutes).
We randomly chose the following three coding issues from the shortlist:
\begin{enumerate}[label=\textbf{T\arabic*.}, ref=T\arabic*, leftmargin=*]
\item \label{enu:user-study:constructors}
  Ensure that constructors do not call overridable methods (MET05-J).
\item \label{enu:user-study:equals}
  Do not use the \lstinline|Object.equals()| method to compare two arrays (EXP02-J).
\item \label{enu:user-study:private}
  Do not return references to private mutable class members (OBJ05-J).
\end{enumerate}


\paragraph{Study structure}

Participants completed the 3 tasks in a counter-balanced within-subject design, where we randomly assigned
each task as either a \emph{treatment} condition (with access to \Tool) or a \emph{control}
condition (without access to \Tool). We ensured that each participant completed at least one task
with access to \Tool{} and at least one task without access to \Tool.
We randomized task order to mitigate potential order effects and other confounding factors.
We informed participants that they could use any available tool to complete tasks. We recorded
screens and audio, and stored modified codebases anonymously on a secure drive for subsequent
analysis, as per our university's IRB guidelines.
We set a 40-minute time limit for each task.
Upon completion,
participants completed a post-study questionnaire regarding their
experience and provided feedback on \Tool.


\paragraph{Data analysis}
We calculated participant accuracy by comparing the number of code locations that they identified to
the actual number of coding issues present, which we obtained using a hand-crafted CodeQL query.
Next, two researchers performed a negotiated agreement to label each of the participant fixes as
either plausible (1) or not (0). We calculated fix accuracy as the average across all three tasks
for each participant.

\mmclearpage \subsection{\ref{enu:user-study:impact-process}: \Tool's impact on answer seeking process}

We identified 3 central themes of how \Tool{} impacted the participants' approach:


\paragraph{Shifting strategies from pattern matching to high level reasoning}

\Tool{} replaced labor intensive pattern
matching with high level reasoning. Without \Tool, most participants relied on literal cues such as
keyword searches or regexes during the finding phase: 15 participants (83\%) used literal matching,
while only 11 (61\%) engaged in object identification strategies.
In contrast, when using \Tool, object identification became dominant: No participants used literal
cues, while 14 participants (77\%) adopted object identification approaches.
Participants attributed this change to \Tool's ability to eliminate tedious low level work. As one
participant explained, \Tool{} removed the need to ``\emph{find all occurrences of something by
using regex and judg[e] one by one, [which] is [a] tedious task},'' enabling them to focus on more
intellectually demanding aspects of debugging. Another participant emphasized the magnitude of this
shift, noting that ``\emph{\Tool{} definitely made that task easy. Doing Task 1 without [\Tool{}]
was basically impossible.}''


\paragraph{Lowering reliance on external AI}

\Tool{} significantly changed how participants relied on external AI assistance, affecting both the
frequency and purpose of AI use. In the control group, participants frequently turned to AI during
the finding phase, using it to generate regex patterns for locating potentially buggy code and, in
some cases, to fully delegate bug identification to AI assistants. ChatGPT was typically used for
pattern generation, while higher-capacity models such as Gemini, Cursor AI, and GitHub Copilot were
preferred for broader code analysis due to their larger context windows~\cite{comanici2025gemini}.
When \Tool{} was available, participants rarely used external AI during the finding phase and
instead restricted AI assistance to code-fixing tasks.
However, reduced reliance on external AI may also introduce new risks: users may over-trust \Tool's
outputs, potentially leading to confirmation bias~\cite{zhou2026cognitive} in which its suggestions are accepted without
sufficient critical scrutiny~\cite{sadra-trust, klayman1995varieties}.


\paragraph{Combining the strengths of LLMs and program analyzers}

When debugging tasks were fully delegated to LLMs or AI code agents, participants frequently
experienced high latency and received incomplete or inaccurate results, despite the convenience of
natural language interaction.
Participants consistently highlighted the \Tool's ability to correctly interpret natural language
intent and to support expressive, criteria-driven searches. This capability was particularly valued
by those accustomed to rigid keyword based workflows, who emphasized the benefit of describing bugs
directly rather than by searching for specific terms.

\mmclearpage \subsection{\ref{enu:user-study:effectiveness}: The Effectiveness of \Tool{} in Finding and Fixing Code Issues}

\paragraph{Time to locate issues}

\begin{figure}
    \centering
    \includegraphics[width=0.75\linewidth]{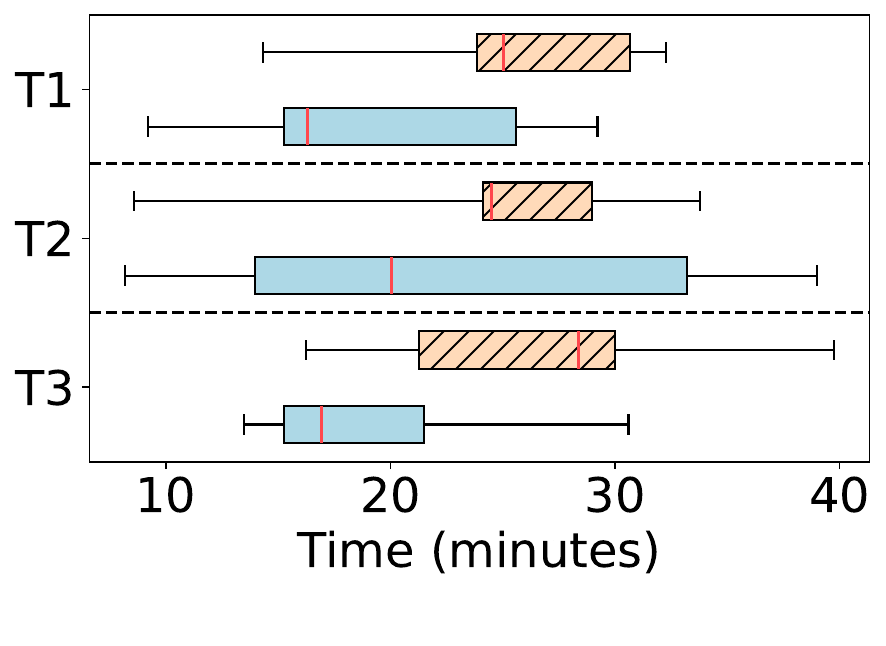}
    \caption{Time needed by participants in the user study.}
    \label{fig:user-study:findAndFix:time}
\end{figure}
We observed that participants using \Tool{} spent 31\% less time on average to complete the tasks
(Figure~\ref{fig:user-study:findAndFix:time}). Indeed, a one-sided $t$-test\footnote{Shapiro-Wilk tests were non-significant ($p > .05$), confirming the normality assumption was met for all tasks spent time.} on the
completion time comparing cases where participants were using \Tool{} showed a significant
reduction for Task~\ref{enu:user-study:constructors} ($t = -1.93$, $p = 0.03$, $ df = 14$) with 35\% reduction
on average and Task~\ref{enu:user-study:private} ($t = -2.12$, $p < 0.01$, $df = 14$) with 40\% reduction on
average. While the average completion time for Task~\ref{enu:user-study:equals} reduced 18\% on
average, we couldn't find a significant difference ($t = -0.06$, $p = 0.47$, $df = 14$). This is unsurprising,
because this task was also the most amenable to a standard \texttt{grep} search.


\paragraph{Accuracy of identification}

Overall, having access to \Tool{} increased task accuracy by an average of $3.8\times$. This
increase in accuracy was greatest in the case of Tasks~\ref{enu:user-study:constructors} and~%
\ref{enu:user-study:private}, where we observed a substantial $5.6\times$ and $4.9\times$
improvement in average accuracy respectively, and least in the case of Task~%
\ref{enu:user-study:equals} where we only observed a $1.6\times$ improvement in the average.
This is unsurprising, because a simple command such as \lstinline|grep -rn '.equals(' .| worked for
most users in the control group. On the other hand, control group participants in Task~%
\ref{enu:user-study:constructors} attempted but mostly failed to construct a suitable \texttt{grep}
query. Some resorted to manually scanning files---an inefficient and ultimately fruitless strategy.
In a unique case, one control group participant even tried to directly use CodeQL but struggled to
formulate a valid query.
Performing a one-sided Wilcoxon signed-rank test\footnote{Shapiro-Wilk tests were significant
($p < .05$) for all task accuracies; therefore, non-parametric tests were used due to
non-normality.} indicated statistically significant improvements in accuracy arising from the use of
\Tool{} for Tasks~\ref{enu:user-study:constructors} ($W = 28$, $p < 0.01$, $r = 0.66$) and~%
\ref{enu:user-study:private} ($W = 34$, $p = 0.01$, $r = 0.63$) but did not show statistical
significance for Task~\ref{enu:user-study:equals} ($W = 6$, $p = 0.125$, $r = 0.38$).

\begin{figure}
    \centering
    \includegraphics[width=0.75\linewidth]{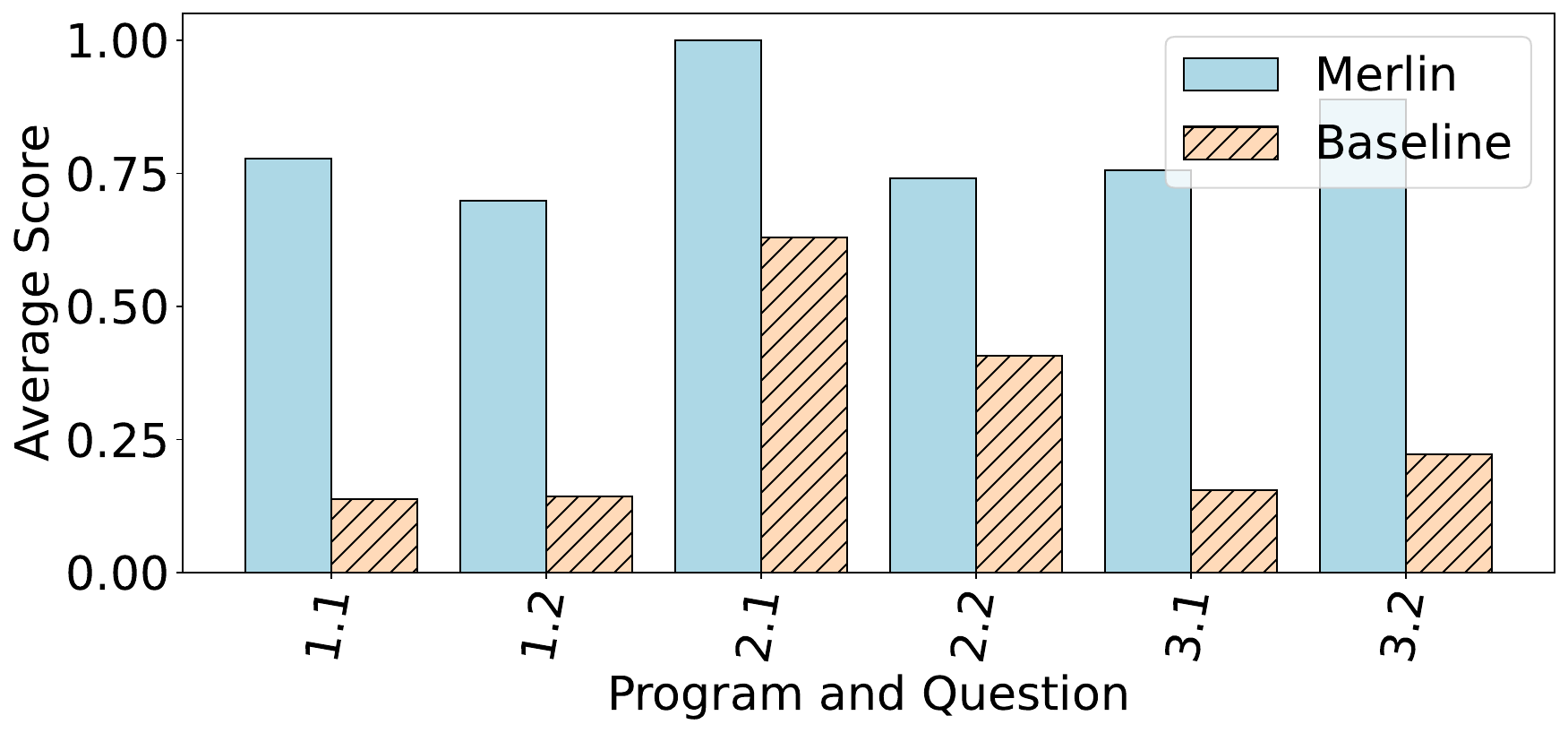}
\caption{Accuracy of responses in the usefulness user study. The measurements for Questions~1.1,~2.1, and~3.1 indicate their accuracy in identifying coding issues, while Questions~1.2,~2.2, and~3.2 measure their effectiveness in fixing them.}
\label{fig:user-study:findAndFix:accuracy}
\end{figure}


\paragraph{Impact on fixing issues}

In Task~\ref{enu:user-study:constructors}, many users addressed the issue by converting overridable
methods that were called within constructors into non-overridable ones by marking them
\lstinline|final|. Experimental group users were more successful than control group users, largely
because \Tool{} had already helped them identify the problematic method calls. Overall, \Tool{} led
to a $4.9\times$ improvement in fix accuracy. Performing a one-sided Wilcoxon signed-rank test revealed a significant
improvement with a large effect size in creating fixes while using \Tool{} ($W = 32$, $p = 0.02$, $r=55$).

In Task~\ref{enu:user-study:equals}, users commonly fixed the issue by replacing expressions such as
\lstinline|array1.equals(array2)| with \lstinline|Arrays.equals(array1, array2)|. However, some
users incorrectly applied this fix to arrays with incompatible types. Despite such errors,
experimental group users showed a $1.8\times$ improvement in accuracy. A one-sided Wilcoxon
signed-rank test revealed a statistically significant improvement with a large effect size
($W = 15$, $p = 0.03$, $r = 0.54$).

Finally, in Task~\ref{enu:user-study:private}, the typical fix involved cloning private mutable
fields before returning them to prevent accidental access to internal state. Some users attempted to
incorrectly clone non-cloneable objects. Overall, experimental group users outperformed the control
group in both accuracy and time, achieving a $4\times$ gain in correctness. However one-sided
Wilcoxon signed-rank test revealed no significant improvement arising from the use of \Tool{}
($W = 24$, $p = 0.06$, $r=0.46$).

\mclearpage \section{Related Work}
\label{sec:related}


\paragraph{Customizable program analyzers}

CodeQL is a ``\emph{semantic code analysis engine}'' that allows programmers to create program
analysis queries using a SQL-like language~\cite{codeql}. It is one of a growing family of such systems,
including SemGrep~\cite{Semgrep} and Amazon CodeGuru / GQL~\cite{GQL}.

Despite their expressiveness, these systems require the user to learn a domain-specific language
(DSL) in order to effectively craft queries. This has motivated the development of
natural language interfaces that are similar to \Tool: Examples include IRIS~%
\cite{Iris} and MoCQ~\cite{MoCQ}. To our knowledge, these systems are either limited in the kinds of
analysis that are supported (e.g., IRIS is tailored to the setting of taint analysis) or require
extensive additional information from the programmer (e.g. MoCQ requires examples of target code and
documentation and samples of the analyzer's DSL). They also target simpler backend analyzers: e.g.,
\citep{MoCQ} acknowledge that CodeQL is a challenging target language and instead primarily focus on
synthesizing queries for Joern~\cite{Joern}, and \citep{Garg2025} target the simpler setting of
structural code search, such as that present in IntelliJ~\cite{JetBrains}, the CodeQue Visual Studio
extension~\cite{CodeQue}, or Comby~\cite{Comboy}. In contrast, \Tool{} supports a larger range of queries and
only requires the question text and schema of the desired output.

A recurring challenge in designing such systems is that the underlying DSL is a low-resource
language necessitating the use of retrieval-augmented generation (RAG) in helping the LLM to produce
the desired analysis query~\cite{RAG}. A notable distinguishing feature of \Tool{} is our automatic
generation of self-tests (instead of user-provided examples in the case of MoCQ), and in the use of
assistive queries to diagnose and fix misbehaving queries. In this context, self-tests are closely
related to the idea of generating unit tests to verify LLM-generated code~\cite{gu2025testartimprovingllmbasedunit, xu2025kodcode}.


\paragraph{Using LLMs to answer questions about code}

Beyond classical transformer-based LLMs~\cite{vaswani2023attentionneed, radford2019language}, more recent agents combine the underlying
statistical model with access to external tools such as \texttt{grep}, \texttt{ls}, and
\texttt{find}. Examples of such systems include Copilot~\cite{Copilot}, Claude Code~\cite{claude-code}, and Cursor~%
\cite{cursor}. Having access to these external tools greatly improves the question-answering capabilities
of these systems, as we observed in Section~\ref{sec:eval}. However, despite using these commands, their
final reasoning process remains expensive, opaque, and non-deterministic~\cite{eibl2025exploring}. In contrast, the final
query executed by \Tool{} serves as a certificate / explanation of its output, and improves the
reliability of the system.

\mclearpage \section{Limitations and Threats to Validity}
\label{sec:discussion}


\paragraph{Threats to validity}

Although we present \Tool{} as a system to answer analytical questions about code, our benchmarks
are only drawn from bug-finding systems. As previously mentioned, this was because research on these
topics serves as a ready source of benchmark questions.

In addition, as discussed in Section~\ref{sec:eval}, we had to manually compose the natural language
questions that described the GitHub Security Lab analyzers. To mitigate possible bias in this
process, two other authors of this paper independently reviewed and edited these translations.

Third, one might be concerned about the small scale of the user study, which consisted of three
tasks and 18 participants of whom 13 were students. We note that the user study of Section~\ref{sec:user-study}
required two entire weeks to schedule and conduct. Performing a larger-scale evaluation with more
participants and a broader range of tasks is a good direction of future work.

Finally, we note that \Tool{} uses gpt-4o as the backend LLM. Therefore, even though we publicly
release our artifact, there is the danger of imperfect reproducibility: This is both because of the
inherent non-determinism of the LLM and its commercial, closed-source nature. Nevertheless, we
expect the broader experimental observations to be stable.


\paragraph{Limitations of our system}
Finally, \Tool{} is not designed to support the full range of developer workflows, such as interactive
debugging or hypothesis-driven exploration, which are known to vary widely across developers and
tasks~\cite{Storey2005, LaToza2010}. Supporting a broader spectrum of workflows is an important
direction for future work.

Our system is currently limited to analyses that the CodeQL backend can perform. Supporting
other program reasoning tools such as symbolic execution engines~\cite{KLEE} and dynamic analysis
frameworks~\cite{Valgrind, AFLpp, ASAN} would greatly extend its capabilities.
%
In addition, the system is currently unable to handle imprecise and vaguely defined predicates
(e.g., ``\emph{untrusted}'' source) and predicates that do not easily permit an exhaustive listing
(e.g., functions which return personally identifiable information).
Extending the system to detect and interactively resolve ambiguous questions~\cite{zhang-choi-2025-clarify} and
supporting non-analytical questions (e.g.,
\guillemotleft{}\emph{Is this code safe?}\guillemotright{} or
\guillemotleft{}\emph{Is this code easy to read?}\guillemotright{}) is an important direction of
future research.

\mclearpage

\bibliography{src/references.bib}

\end{document}